\documentclass[11pt]{article}
\pdfoutput=1

\usepackage{amsmath,amssymb}
\usepackage{epsfig}
\usepackage[numbers,compress,sort]{natbib}
\usepackage{hyperref}
\usepackage{color}
\usepackage{slashed}
\usepackage{placeins}

\usepackage[font={small}]{caption}   

\def\Dslash{\slashed D}

\def\SO{\textrm{SO}}
\def\SU{\textrm{SU}}
\def\U{\textrm{U}}

\definecolor{myred}{rgb}{0.7, 0, 0}
\definecolor{myblue}{rgb}{0, 0, 0.7}
\definecolor{mygreen}{rgb}{0.04, 0.7, 0.5}

\hypersetup{colorlinks,citecolor=myred,linkcolor=myblue,urlcolor=myblue,linktocpage=true}

 \def\be   {\begin{equation}}   \def\ee   {\end{equation}}
 \def\ba   {\begin{array}}      \def\ea   {\end{array}}
 \def\bea  {\begin{eqnarray}}   \def\eea  {\end{eqnarray}}
 \def\bean {\begin{eqnarray*}}  \def\eean {\end{eqnarray*}}

\newcommand{\hhref}[1]{\href{http://arxiv.org/abs/#1}{arXiv:#1}}

\numberwithin{equation}{section}

\title{
\vspace{-2cm}
\begin{flushright}
\small{PREPRINT: DESY 15-243, DFPD-2015-TH-29}
\end{flushright}
\vspace{3cm}
\bf \LARGE
Top Partners Searches and Composite Higgs Models
\vspace{.2cm}}
\date{}
\author{
{\large Oleksii Matsedonskyi$^a$, Giuliano Panico$^{b}$, Andrea Wulzer$^{c}$}\\
[10mm]
\normalsize\itshape $^a$ DESY, Notkestrasse 85, 22607 Hamburg, Germany\\
\normalsize\itshape $^b$ IFAE, Universitat Aut\` onoma de Barcelona, E-08193 Bellaterra, Barcelona, Spain\\
\normalsize\itshape $^c$ Dipartimento di Fisica e Astronomia and INFN, Sezione di Padova,\\
\normalsize\itshape Via Marzolo 8, I-35131 Padova, Italy\\
}

\begin{document}
\maketitle
\begin{abstract}
\medskip
\noindent
Colored fermionic partners of the top quark are well-known signatures of the Composite Higgs scenario and for this reason they have been and will be subject of an intensive experimental study at the LHC. Performing an assessment of the theoretical implications of this experimental effort is the goal of the present paper. We proceed by analyzing a set of simple benchmark models, characterized by simple two-dimensional parameter spaces where the results of the searches are conveniently visualized and their impact quantified. We only draw exclusion contours, in the hypothesis of no signal, but of course our formalism could equally well be used to report discoveries in a theoretically useful format.
\end{abstract}

\newpage

\tableofcontents

\newpage


\section{Introduction}

After several years of thoughtful investigation, the generic idea of the Higgs boson being composite at the TeV scale, addressing the Naturalness Problem associated with its mass, converged to a rather specific framework, with rather specific assumptions, which we denote as ``Composite Higgs'' (CH) scenario. These assumptions, extensively reviewed in refs.~\cite{Contino:2010rs,Panico:2015jxa}\footnote{See ref.~\cite{Bellazzini:2014yua} and references therein for an overview of alternative constructions with a composite Higgs.}, include the fact that the Higgs is a pseudo-Nambu--Goldstone boson (pNGB) \cite{Kaplan:1983fs} (possibly but not necessarily associated with the minimal symmetry breaking pattern \mbox{SO$(5)\rightarrow$ SO$(4)$} \cite{Agashe:2004rs}, which we will assume here) and the generation of fermion masses through the mechanism of partial compositeness \cite{Kaplan:1991dc}. It is this latter hypothesis that makes composite partners of the SM fermions appear in the theory, and in particular the top partners that are the subject of the present study. Actually, partial compositeness can be argued to be strictly needed in the top quark sector only, while alternative mechanisms based on bilinear fermion couplings to the composite sector (as opposite to the linear couplings in partial compositeness) can be considered for the generation of the light quarks and leptons masses \cite{Matsedonskyi:2014iha,Cacciapaglia:2015dsa}. The analysis of the present paper is largely insensitive to the structure of light quarks and lepton couplings because in most scenarios these couplings are too weak to contribute to the top partner's collider phenomenology. Notable exceptions are flavor-symmetric \mbox{$\U(3)^3$} models~\cite{Redi:2011zi} and (to lesser extent) the constructions based on \mbox{$\U(2)^3$} flavor group~\cite{Barbieri:2012uh}, which predict additional sizable signals to be investigated separately~\cite{Delaunay:2010dw}.

The existence of the top partners, {\it{i.e.}} colored fermionic resonances with TeV-sized mass coupled to top and bottom quarks, is an unavoidable universal prediction of partial compositeness in the top sector. The Electro-Weak (EW) quantum numbers of the top partners, their (single) production rate and their decay modes, thus in turn their experimental signatures, are instead model-dependent. Because of this, setting up a comprehensive top partner search program at the LHC and drawing its theoretical implications on the CH scenario results in a non-trivial task. Several aspects of this problem have been addressed and substantially solved in the literature. In particular, some of the most generic production and decay channels of the top partners were identified and studied already in refs.~\cite{Contino:2006qr,Contino:2008hi,AguilarSaavedra:2009es,Mrazek:2009yu} and the analysis was completed and systematized in ref.~\cite{DeSimone:2012fs}.\footnote{Top partners have some similarities with the so-called ``Vector-Like Quarks'' (VLQ) \cite{delAguila:2000aa,delAguila:2000rc,AguilarSaavedra:2009es,Aguilar-Saavedra:2013qpa}, but also radical differences. VLQ's are described by renormalizable Lagrangians and couple to quarks through mass-mixings induced by Yukawa couplings. Top partners possess non-renormalizable interactions that are dictated by the pNGB nature of the Higgs and have important implications on their mass spectrum and on their couplings. The reduction of the charge-$5/3$ VLQ single production rate, which is instead considerable for the top partners, is one example of these differences which we will discuss in section~\ref{l4p}. Also related, but different, are the ``top partners'' encountered in Little Higgs theories \cite{Perelstein:2003wd,Han:2003wu,Berger:2012ec,Kearney:2013cca}, whose origin, nature and properties are dictated by the additional Higgs mass protection mechanism \cite{ArkaniHamed:2002qy} that is present in those theories.} More exotic possibilities, also including the possible interplay with other resonances of the composite sector, have been considered in refs.~\cite{Bini:2011zb,Matsedonskyi:2014lla,Greco:2014aza,Chala:2013ega,Chala:2014mma,Azatov:2015xqa,Araque:2015cna}. As a result of this work, a number of final states and signal topologies have been identified where to search for top partners at the LHC. 

The second aspect of the top partner phenomenology that has been largely addressed in the literature concerns the complete experimental exploration of the possible top partner signals and the presentation of the search results in a meaningful and useful way. When restricting to the QCD pair-production mode, a valid strategy is the one adopted by ATLAS and CMS in the $8$~TeV analyses, which provide combined mass limits as a function of the top partners branching ratios in the allowed SM decay channels. Alternative strategies will have to be adopted to deal with single-production in the $13$~TeV analyses. Several proposals have been made, ranging from the usage of explicit models~\cite{Aguilar-Saavedra:2013qpa} to the implementation of an automatic recasting tool~\cite{Buchkremer:2013bha,Barducci:2014ila}. The strategy we proposed in ref.~\cite{Matsedonskyi:2014mna}, which we will employ in the present paper, consists in reporting the result of each search in the appropriate mass-coupling plane of a phenomenological Lagrangian, suited for being easily interpreted in more complete explicit models.

The third aspect of the problem, on which we aim to make progresses here, is how to draw the theoretical implication on the CH scenario of all this (past and future) work. Namely we would like to quantify what current top partner exclusions are telling us on the CH idea and what would come from future exclusions or, more optimistically, from future discoveries. In view of the above-mentioned model-dependence, explicit benchmark models are needed for this purpose (and for this purpose only). Those benchmarks have to be judiciously chosen to be representative of a wide class of theoretical possibilities. The logic by which we perform this choice is explained below.

Top partners are important in CH because they are connected with the generation of the Higgs potential and thus in turn with the physical Higgs boson mass and with the EW Symmetry Breaking (EWSB) scale $v\simeq246$~GeV. This can be seen to imply that in basically all CH models (interesting exceptions are discussed in the Conclusions) the top partners have to be rather light for the Higgs being as light as observed ($m_H\simeq125$~GeV) and the theory being ``Natural''. Namely, the top partners mass is related with the level of fine-tuning $\Delta$ in the theory, which is the essential parameter to be kept under control in those models, like CH, whose {\it{raison d'\^etre}} is addressing the Naturalness Problem. The relation reads $\Delta\geq (M_{\textrm{Partner}}/{450\,{\textrm{GeV}}})^2$. While this lower bound on $\Delta$ holds in general, the actual form of the mass/tuning connection and how it is influenced by the other parameters depends on how the partial compositeness hypothesis is implemented in the theory and the Higgs potential emerges~\cite{Matsedonskyi:2012ym,Panico:2012uw}. The two classes of models that we will consider, which we dub ``minimal tuning'' and ``double tuning'' scenarios using the terminology of ref.~\cite{Panico:2012uw}, correspond to the two known types of implementations.\footnote{In spite of the names, double tuning models are not generically more fine tuned than the minimal tuning ones, at least with the definition of fine-tuning given below. The name refers to the amount of tuning needed to adjust the EWSB scale.}

The structural differences between the two scenarios and the benchmark models they lead us to will be described in the following sections. Here we instead outline their common features and specify our definition of fine-tuning, which we obviously take to be the same in the two cases for a fair comparison. In both our scenarios, the Higgs potential takes the form~\cite{Panico:2015jxa}
\be
\displaystyle
V[H]=-\alpha f^2\sin^2\frac{|H|}{f}+\beta f^2\sin^4\frac{|H|}{f}\,,
\ee
where $f$ is the Higgs decay constant, {\it{i.e.}} the order parameter for the breaking of \mbox{SO$(5)\rightarrow$ SO$(4)$}, and $\alpha$ and $\beta$ are model-dependent radiatively generated coefficients. In order to obtain the correct Higgs mass and EWSB scale, $\alpha$ and $\beta$ need to assume the values
\be
\label{parneeded}
\displaystyle
\alpha=\alpha_{\textrm{needed}}=\frac{m_H^2}{4(1-\xi)}\,,\;\;\;\;\;\beta=\beta_{\textrm{needed}}=\frac{\alpha_{\textrm{needed}}}{2\xi}\,,
\ee
where $\xi=v^2/f^2$ is the famous CH parameter that controls all the departures from the SM. In particular, $\xi$ controls the modifications of the Higgs boson couplings and it is bounded to $\xi\lesssim0.1$ at $95\%$~CL
by Higgs coupling measurements~\cite{Aad:2015pla}.

The needed values of $\alpha$ and $\beta$ in eq.~(\ref{parneeded}) should be compared with the expected ``Natural'' size of these parameters: $\alpha_{\textrm{expected}}$ and $\beta_{\textrm{expected}}$. If they are much smaller, engineering them requires large cancellations of order
\be
\displaystyle
\Delta_\alpha=\frac{\alpha_{\textrm{expected}}}{\alpha_{\textrm{needed}}}\,,\;\;\;\;\;\Delta_\beta=\frac{\beta_{\textrm{expected}}}{\beta_{\textrm{needed}}}\,.
\ee
The minimal and double tuning scenarios produce different estimates of the expected $\alpha$ and $\beta$ and thus in turn different estimates for $\Delta_\alpha$ and $\Delta_\beta$. Actually, a universal formula that holds in the two cases can be written for $\alpha_{\textrm{expected}}$ and thus in turn for $\Delta_\alpha$. It is
\be
\label{tuning}
\displaystyle
\alpha_{\textrm{expected}}=\frac{3}{16\pi^2}\lambda_t^2 M_\Psi^2\,,\;\;\;\Rightarrow\;\;\;\Delta_\alpha\simeq\frac{3}{4\pi^2}\lambda_t^2 \frac{M_\Psi^2}{m_H^2}\simeq\lambda_t^2\left(\frac{M_\Psi}{450\,{\textrm{GeV}}}\right)^2\,,
\ee
where $\lambda_t$ is a parameter that sets the strength of the top quark interaction with the composite sector and controls, among other things, the generation of the top mass. The size of $\lambda_t$ is related with the top Yukawa, $y_t$, but the relation is different in the minimal and double tuning scenarios leading, as we will see, to different fine-tuning estimates in the two cases. In both cases, instead, $M_\Psi$ is the typical top partners mass scale (not necessarily the mass of the lightest top partner resonance). It sets the energy scale of $\alpha$, {\it{i.e.}} the one of the Higgs mass-term, because it corresponds to the scale where the Higgs potential is generated or, poorly speaking, the one at which the top loop quadratic divergence is canceled. Clearly, $M_\Psi$ is bounded from below by the mass of the lightest top partner state. Furthermore, it turns out that $\lambda_t$ cannot be smaller than the top Yukawa $y_t\simeq1$ if we want to generate the correct top mass. Therefore $\lambda_t\geq y_t\simeq1$ in all scenarios. These lower bounds produce the above-mentioned universal relation between the top partner mass and the minimal allowed level of tuning. The actual tuning, which we will estimate by applying eq.~(\ref{tuning}) with the value of $\lambda_t$ which is appropriate in each model, can be larger and thus the mass/tuning connection can be stronger.

The second parameter in the potential, $\beta$, originates in a radically different way in the minimal and double tuning cases so that its expected size can no be cast in a universal formula. However this is not a problem because the cancellation of $\beta$, which is required  in some regions of the parameter space, needs not to be taken into account in the definition of the total level of fine-tuning $\Delta$. More precisely, it needs not to be taken into account if the tuning is defined, following the philosophy of ref.~\cite{Barbieri:1987fn}, as the maximal amount of cancellation taking place in the theory.\footnote{Alternative definitions might require, for instance, to multiply $\Delta_\alpha$ with $\Delta_\beta$.} This is because $\Delta_\alpha$ is systematically larger than $\Delta_\beta$, a fact that can be easily established by observing that $\alpha_{\textrm{expected}}$ is either larger than $\beta_{\textrm{expected}}$ (in the double tuning case), or comparable (minimal tuning)~ \cite{Panico:2012uw}. Therefore, using eq.~(\ref{parneeded})
\be
\Delta_\beta=\frac{\beta_{\textrm{expected}}}{\alpha_{\textrm{expected}}}\frac{\alpha_{\textrm{expected}}}
{\beta_{\textrm{needed}}}=2\,\xi \frac{\beta_{\textrm{expected}}}{\alpha_{\textrm{expected}}}\Delta_\alpha< \Delta_\alpha.
\ee
The total tuning is provided by the largest cancellation, thus we set $\Delta=\Delta_\alpha$.

The rest of the paper is organized as follows. In Section~\ref{sec:MT} and~\ref{sec:DT} the minimal and double tuning scenarios are  discussed separately and the corresponding benchmark models are defined and analyzed. The impact of current top partners exclusions and the projections for the $13$~TeV run are quantified for each benchmark. Combined with the tuning estimate from eq.~(\ref{tuning}), this allows us to estimate how much the Natural parameter space region of the CH scenario has been excluded by the $8$~TeV run and how much of it will be tested by the forthcoming one. Limits are obtained by the procedure of ref.~\cite{Matsedonskyi:2014mna}, whose implementation is described in some detail in Appendix~\ref{sec:bounds}. In the Appendix we also present a reassessment of the current and future experimental situation in view of recent studies on top partners collider searches. Finally, we present our Conclusions on what the LHC could tell us about the CH idea and on new model-building directions it could push us towards.

\section{Minimal Tuning: \boldmath$\mathbf{14 + 1}$}\label{sec:MT}

As a first class of models we consider the ones that represent the ``minimal tuning'' case. A set-up realizing this type of theories is
obtained by assuming that the $\SU(2)_L$ doublet $q_L = (t_L, b_L)$, following the partial compositeness assumption,
is linearly mixed with composite operators transforming in the $\mathbf{14}$ representation of $\SO(5)$.
The right-handed $t_R$ component, on the other hand, can be either mixed with composite operators that are singlets
under $\SO(5)$, or realized as a composite chiral singlet originating directly from the strongly-coupled dynamics.
This set-up is usually denoted as the $14 + 1$ scenario~\cite{Panico:2012uw,DeSimone:2012fs}.
The amount of tuning is minimized if the $t_R$ field is a fully composite state or
an elementary state with a large, nearly maximal coupling with the composite dynamics. In both cases the phenomenology
of the model is quite similar, the only difference being a minor modification in the estimates for the coefficients
in the effective Lagrangian. For definiteness in the following we will concentrate on the scenario with a fully composite $t_R$
and we will only briefly comment on the differences that arise in the partially composite case.

From the decomposition of the $\mathbf{14}$ representation under $\SO(4)$, namely
\begin{equation}
\mathbf{14} = \mathbf{9} \oplus \mathbf{4} \oplus \mathbf{1}\,,
\end{equation}
we infer that the top partners in the $14 + 1$ scenario must fill nineplet, fourplet or singlet representations
of the unbroken $\SO(4)$ subgroup. For our purposes it is convenient to consider only the lightest composite partners, which are
the ones that most directly affect the collider phenomenology and are most easily accessible in direct searches.
We will thus focus on simplified scenarios in which
only one $\SO(4)$ multiplet of fermionic partners is light, while the others are heavy enough so that their contributions
can be safely neglected.\footnote{Notice that this assumption is not particularly restrictive. Given the steep fall of the
parton distribution functions, mass differences of a few hundred GeV between the heavier states and the lightest partners
are already enough to ensure that the collider phenomenology is completely dominated by the lightest resonances.}
The set-up with a light $9$-plet has been thoroughly analyzed
in ref.~\cite{Matsedonskyi:2014lla}, where a bound $m_9 \geq 990\ \mathrm{GeV}$ has been derived on the mass
of the multiplet by using the $8\ \mathrm{TeV}$ LHC data. In the $14\ \mathrm{TeV}$ LHC run the bound is expected to
reach $m_9 \gtrsim 1.9\ \mathrm{TeV}$ for an integrated luminosity ${\mathcal L} \simeq 100\ \mathrm{fb}^{-1}$.
In the following we will thus focus only on the scenarios characterized by a light $4$-plet or a light singlet.

\subsection{Light fourplet}
\label{l4p}

The most general leading-order effective action for a light fourplet $\psi_4$ can be easily written by using the
CCWZ framework~\cite{ccwz}
\begin{eqnarray}
{\cal L} &=&  i\, \overline q_L \Dslash q_L + i\, \overline t_R \Dslash t_R + i\overline \psi_4 (\Dslash - i \slashed e) \psi_4
- m_4 \overline \psi_4 \psi_4
\nonumber\\
&& + \left(-i\, c_t \overline \psi_{4R}^i \gamma^\mu d_\mu^i t_R
+ \frac{y_{Lt}}{2} f (U^t \overline q_L^{\bf 14} U)_{55} t_R +
y_{L4} f (U^t \overline q_L^{\bf 14} U)_{i5} \psi_4^i + {\rm h.c.}\right)\,.
\label{eq:fourplet_14+1}
\end{eqnarray}
For an in-depth explanation of the formalism and for the detailed definitions of the notation we refer the reader to
ref.~\cite{Panico:2015jxa}.\footnote{Our notation can be easily matched with the one of ref.~\cite{DeSimone:2012fs},
namely $y_{Lt} \equiv y$, $y_{L4} \equiv y\, c_2$ and $c_t \equiv c_1$.}
Here we only include a brief definition of the main objects.
The embedding of the $q_L$ doublet into the representation $\mathbf{14}$ is denoted by $q_L^{\bf 14}$
and its explicit form is
\begin{equation}\label{eq:embedding_14+1}
q_L^{\mathbf{14}} = \frac{1}{\sqrt{2}}
\left(
\begin{array}{ccccc}
0 & 0 & 0 & 0 & -i\,b_L\\
0 & 0 & 0 & 0 & -b_L\\
0 & 0 & 0 & 0 & -i\,t_L\\
0 & 0 & 0 & 0 & t_L\\
-i\, b_L & -b_L & -i\,t_L & t_L & 0
\end{array}
\right)\,.
\end{equation}
The four Goldstone components, which are identified with the Higgs multiplet $\Pi_i$, in the real fourplet
notation, are described by the matrix
\begin{equation}
U \equiv \exp\left[i \frac{\sqrt{2}}{f} \Pi_i \widehat T^i\right]\,,
\end{equation}
where $\widehat T^i$ $(i=1, \ldots, 4)$ are the generators of the $\SO(5)/\SO(4)$ coset
and $f$ is the Goldstone decay constant. On the first line of eq.~(\ref{eq:fourplet_14+1}), $D_\mu$ denotes the
standard covariant derivative containing the SM elementary gauge fields.
Finally the $d_\mu$ and $e_\mu$ objects denote the CCWZ operators, which can be defined in terms
of the Maurer--Cartan form constructed from $U$, namely
\begin{equation}\label{eq:d_e_symbol}
U^t[A_\mu + i\partial_\mu]U \equiv d_\mu^i \widehat T^i + e_\mu^a T^a\,,
\end{equation}
where $T^a$ $(a=1, \ldots,6)$ denote the $\SO(4)$ generators. In eq.~(\ref{eq:d_e_symbol}) $A_\mu$ corresponds to
the SM gauge fields rewritten in an $\SO(5)$ notation
\begin{equation}\label{eq:gauge}
A_\mu = \frac{g}{\sqrt{2}} W^+_\mu (T^1_L + i T^2_L) + \frac{g}{\sqrt{2}} W^-_\mu (T^1_L - i T^2_L)
+ g (c_w Z_\mu + s_w A_\mu) T^3_L + g'(c_w A_\mu - s_w Z_\mu) T^3_R\,.
\end{equation}
where $g$ and $g'$ are the couplings of the $\SU(2)_L$ and $\U(1)_Y$ subgroups
and $c_w$, $s_w$ are the cosine and sine of the weak mixing angle, $\tan \theta_w = g'/g$.

To complete the description of the effective parametrization it is useful to discuss the power-counting associated to the
parameters in the Lagrangian~\cite{Giudice:2007fh,Panico:2015jxa}.
Since we focused on the scenario in which the $t_R$ field is fully composite,
the $d_\mu$-symbol interaction fully arises from the composite dynamics, the corresponding coefficient $c_t$ is thus
expected to be of order one. The other operators on the second line of eq.~(\ref{eq:fourplet_14+1}), on the other hand,
involve an elementary and a composite field, thus they their size is not dictated by the
composite dynamics but it depends on the elementary/composite interaction strength in the UV theory.
On general grounds we expect all the interactions of the $q_L$ doublet with the composite sector to originate
from a single dominant operator in the UV, thus implying $y_{Lt} \sim y_{L4}$.
The size of $y_{Lt}$ is then fixed by the requirement of reproducing the correct top mass.
Let us now briefly discuss how the above estimates are modified if we assume that the $t_R$ is not a fully composite state,
but instead is associated to an elementary field. In this case the $d_\mu$-symbol interaction does not fully arise from
the composite sector, thus it is expected to be weighted by an elementary/composite mixing $y_R$.
Its coefficient can be estimated as $c_t y_R/g_*$, where $g_*$ denotes the typical composite sector coupling.
Analogously the operator involving the $t_R$ and $q_L$ fields has now a natural coefficient $y_{Lt} y_R/g_*$.
As expected, in the limit of large elementary/composite mixing $y_R \sim g_*$, the modified estimates give back the
results for a fully composite $t_R$.

We can now describe the features of the spectrum of the fermionic states.
The top mass is mainly determined by the direct mass term in the effective Lagrangian and is controlled by the parameter
$y_{Lt}$. Neglecting higher-order terms in the $v/f$ expansion, we find the following expression
\begin{equation}
m_{top}^2 = \frac{1}{2} \frac{m_4^2}{m_4^2 + y_{L4}^2 f^2} y_{Lt}^2 f^2 \xi\,.
\end{equation}
Notice that the EWSB scale $v\sim246$~GeV (as set by the $W$ mass formula or by the Fermi constant) does not coincide with the Higgs field VEV, but is related to the latter by
\begin{equation}
v^2=f^2(\sin{\langle{\Pi_4}\rangle/f})^2= f^2\xi \,.
\end{equation} 
In addition to the top, the spectrum contains $4$ composite fermionic resonances coming from the
$\psi_4$ multiplet. It is convenient to decompose $\psi_4$ in fields with definite quantum numbers under the SM group:
\begin{equation}\label{eq:4-plet_structure}
\psi_4 = \frac{1}{\sqrt{2}}
\left(
\begin{array}{c}
-i B + i\, X_{5/3}\\
-B - X_{5/3}\\
-i\,\hat T - i\,\hat X_{2/3}\\
\hat T - \hat X_{2/3}
\end{array}
\right)\,.
\end{equation}
The four components correspond to two $\SU(2)_L$ doublets, $(\hat T, B)$ and $(X_{5/3}, \hat X_{2/3})$, with
hypercharges $1/6$ and $7/6$ respectively. The first doublet has the same quantum numbers of the
elementary $q_L$ doublet, while the second one contains an exotic state, the $X_{5/3}$
with electric charge $5/3$ and a top-like state, the $\hat X_{2/3}$ with charge $2/3$.
It can be easily checked that one combination of the $\hat T$ and $\hat X_{2/3}$ states,
which we will denote by $X_{2/3}$, has no mass mixing with the other fields, thus its
mass is just given by $m_{X_{2/3}} = m_4$. This state is degenerate in mass with the $X_{5/3}$ resonance,
whose exotic charge prevents any mixing with the other fermionic states.
The remaining charge-$2/3$ resonance, which we will denote by $T$, is mixed with the $t_L$ field and
receives an additional mass shift after EWSB. Its mass, up to corrections of higher-order in $v/f$, is given by
\begin{eqnarray}
m_T &\simeq& \sqrt{m_4^2 + y_{L4}^2 f^2}\left[1 - \frac{5 y_{L4}^2 f^2}{4 m_4^2}\xi + \cdots\right]\,,
\end{eqnarray}
where inside the square brackets we only kept the leading order terms in an expansion
in the elementary/composite mixings.\footnote{The identification of the heavy mass eigenstates with composite resonance fields is only valid as long as the elementary/composite mass-mixings are much smaller
than the mass parameters in the composite sector. Otherwise the eigenstates develop a significant component along the elementary degrees of freedom and/or the $t_R$.}

\begin{figure}
\centering
\includegraphics[width=0.3\textwidth]{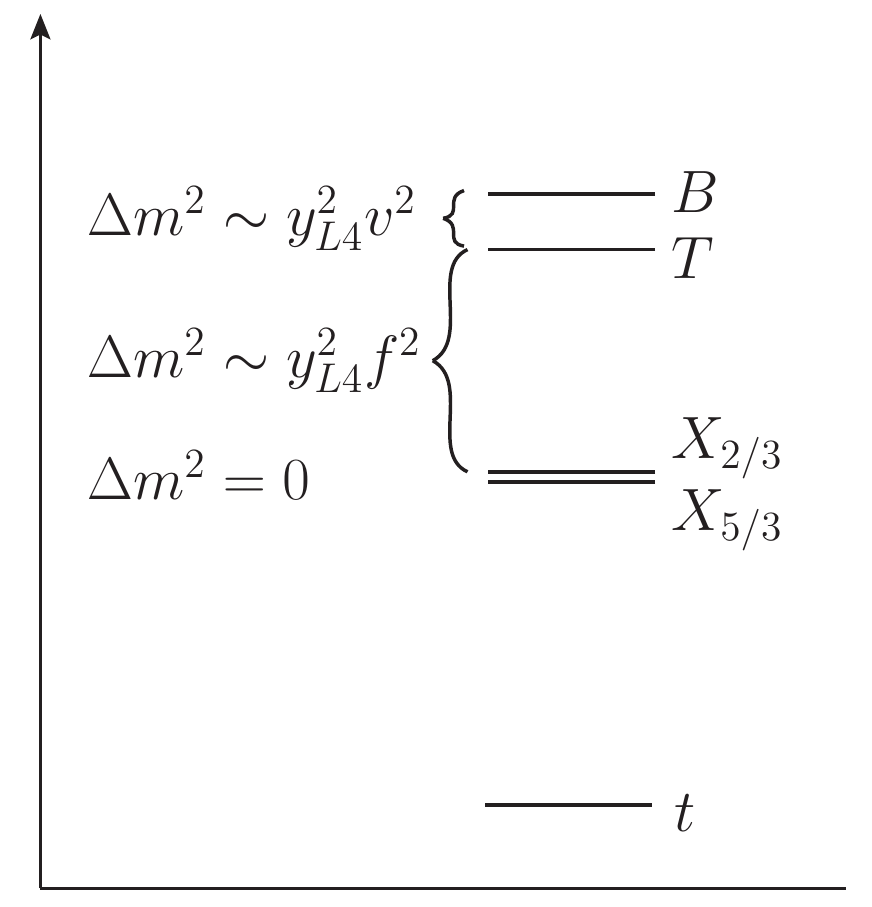}
\caption{Typical mass spectrum of the fourplet states in the $14 + 1$ model.}
\label{fig:mass_spectrum_comptR}
\end{figure}

The fermionc spectrum is completed by a charge $-1/3$ field, the $B$, which is mixed with the $b_L$ component
and whose mass reads
\begin{equation}
m_B = \sqrt{m_4^2 + y_{L4}^2 f^2 (1 - \xi)}
\simeq \sqrt{m_4^2 + y_{L4}^2 f^2} \left[1 - \frac{y_{L4}^2 f^2}{2 (m_4^2 + y_{L4}^2 f^2)}\xi + \cdots\right]\,.
\end{equation}
Together with the $T$ resonance, the $B$ field forms a nearly degenerate doublet.
The overall structure of the spectrum of the quadruplet fields is shown in fig.~\ref{fig:mass_spectrum_comptR}.

Notice that in the effective Lagrangian we did not include a right-handed bottom component,
therefore the $b_L$ remains in the spectrum as a massless field.
This choice is motivated by the fact that the mixing of the $b_R$ quark to the composite dynamics
is typically much smaller than the one of the top, due to the smaller value of the bottom mass. For this reason
the $b_R$ field has only a marginal impact on the collider phenomenology. We will adopt the same simplification
also in the other models analyzed in the following sections.

We can now discuss the impact of the experimental searches on the parameter space of this benchmark model.
For this aim we follow the procedure of ref.~\cite{Matsedonskyi:2014mna}, which we will briefly summarize
in Appendix~\ref{sec:bounds}. The model has a total of three free parameters, given that one of the parameters
in the Lagrangian has to be fixed in oder to the reproduce the correct top quark mass. We decided to fix $y_{Lt}$ because
it is the one to which the top mass shows the largest sensitivity and we are left with the three free quantities $m_4$, $y_{L4}$
and $c_t$.\footnote{The value of $y_{Lt}$ that reproduces the top mass 
depends very mildly on the other parameters and in a large part of the parameter space almost coincides with the top
Yukawa $y_{Lt} \simeq y_{top}$.
For our numerical analysis we fix the top mass to the running value $m^{(2\ \mathrm{TeV})}_{top} = 150\ \mathrm{GeV}$,
which corresponds to a pole mass $m_{top} = 173\ \mathrm{GeV}$.}

\begin{figure}
\centering
\includegraphics[height=.31\textwidth]{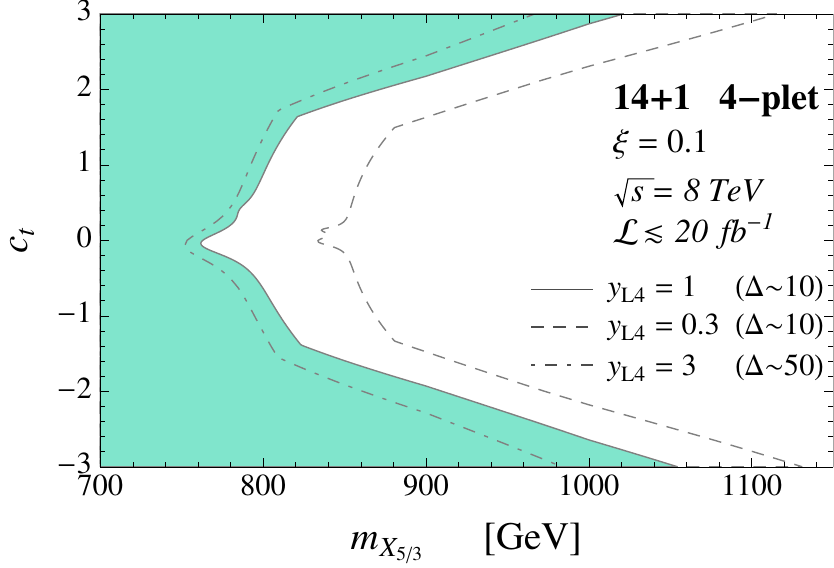}
\hspace{1em}
\includegraphics[height=.304\textwidth]{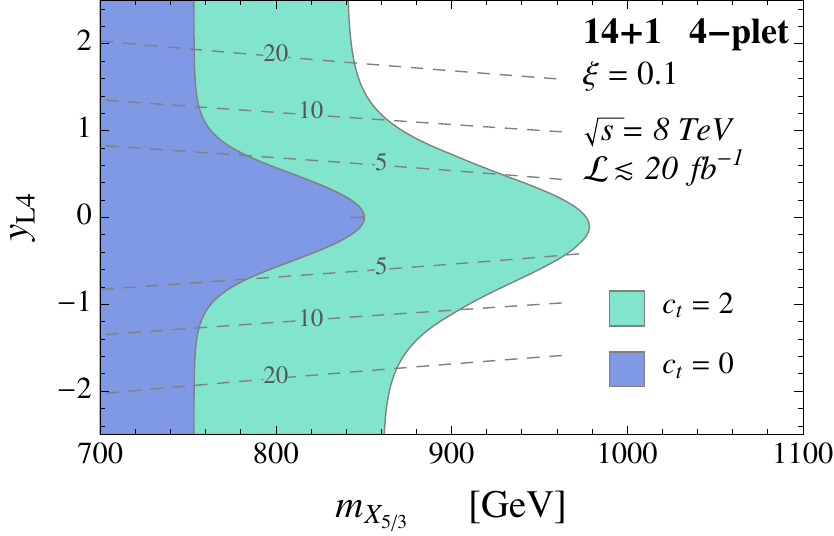}
\caption{Exclusions in the $14 + 1$ model with a light composite fourplet for the $8\ \mathrm{TeV}$ LHC data.
On the left panel: exclusions in the $(m_{X_{5/3}}, c_t)$ plane for $\xi = 0.1$.
The green region shows the excluded points for $y_{L4} = 1$, while the dot-dashed and dashed lines
correspond to $y_{L4} = 3$ and $y_{L4} = 0.3$ respectively. The approximate amount of tuning $\Delta$
associated to each value of $y_{L4}$ for $m_{X_{5/3}} \sim 1\ \mathrm{TeV}$ is given in the legend.
On the right panel: exclusions in the $(m_{X_{5/3}}, y_{L4})$ plane.
The blue (green) region shows the excluded points for $c_{t} = 0$ ($c_{t} = 2$) for $\xi = 0.1$.
The dashed lines show the amount of tuning $\Delta$ computed by using eq.~(\ref{eq:tuning_14+1_4plet}).
}
\label{fig:4plet-14+1_8TeV}
\end{figure}

In fig.~\ref{fig:4plet-14+1_8TeV}, we plot the exclusion bounds obtained from the run 1 LHC data.
The results are presented in the planes $(m_{X_{5/3}} = m_4, c_t)$ and $(m_{X_{5/3}} = m_4, y_{L4})$.
For illustrative purposes we fixed
the Goldstone decay constant $f$ to the value $f = 780\ \mathrm{GeV}$, which corresponds to $\xi = 0.1$.
This value coincide with the present exclusion bound coming from the Higgs coupling measurements~\cite{Aad:2015pla},
and is also suggested by the compatibility with the EW precision data~\cite{Grojean:2013qca}. Limit projections at $13$~TeV, obtained by the rough extrapolation procedure outlined in ref.~\cite{Matsedonskyi:2014mna} and in Appendix~\ref{sec:bounds}, are displayed in fig.~\ref{fig:4plet-14+1_13TeV} in the $(m_{X_{5/3}} = m_4, c_t)$ plane. The integrated luminosity is fixed to $20$~fb$^{-1}$ on the left panel and three curves are shown at different values of $y_{L4}$, while on the right panel $y_{L4}=1$ and the integrated luminosity ranges from $100$~fb$^{-1}$ to $3$~ab$^{-1}$.

The impact of the two parameters $c_t$ and $y_{L4}$ on the bounds is quite easy to understand.
At leading order, the relevant $X_{5/3}$ coupling is independent of $y_{L4}$ and just scales linearly with $c_t$
and with $v/f$,
\begin{equation}\label{eq:coupl_X53}
g_{X_{5/3}t_R} = \frac{g}{2} c_R^{X_{5/3}W} = \frac{g}{\sqrt{2}} c_t \frac{v}{f}\,.
\end{equation}
In the above formula the $c_R^{X_{5/3}W}$ parametrization is included to make contact with ref.~\cite{Matsedonskyi:2014mna},
whose procedure and results we used to derive the bounds in the present analysis.
As can be seen from eq.~(\ref{eq:coupl_X53}), a larger value of $c_t$ enhances the single production channel and tightens
the bounds. The $y_{L4}$ parameter, instead, has an indirect effect on the exclusions since it determines the mass split between
the $B$ resonance and the $X_{5/3}$,
$\Delta m^2 \sim y_{L4}^2 f^2$. At small values of $y_{L4}$ the two states are nearly degenerate
and, since they contribute to the same final state, the signal is enhanced.
At large values of $y_{L4}$ the $B$ gets much heavier and its contribution to the signal becomes negligible.
In this situation the bounds are only driven by the $X_{5/3}$ signal.

Let us now turn to the estimate of the level of fine-tuning of the model. We apply eq.~(\ref{tuning}), in which ``$\lambda_t$''
should be interpreted as the strength of the elementary/composite top sector couplings that break the Goldstone symmetry and
thus generate the Higgs potential. The parameters that break the Goldstone symmetry in our model are $y_{Lt}$ and $y_{L4}$,
therefore the tuning estimate reads
\begin{equation}\label{eq:tuning_14+1_4plet}
\Delta \simeq (y_{L4}^2 + y_{Lt}^2) \left(\frac{m_4}{450\ \mathrm{GeV}}\right)^2\,.
\end{equation}
Notice that $\Delta$ is independent of the value of $c_t$, since this coupling parametrizes
a purely composite-sector operator that is invariant under the Goldstone symmetry and does not contribute
to the Higgs potential. Contour lines of $\Delta$ obtained with the above formula are reported in the right panels of fig.~\ref{fig:4plet-14+1_8TeV} (exploiting the fact that $\Delta$ is independent of $c_t$) and of fig.~\ref{fig:4plet-14+1_13TeV}. On the left panels, instead, the $\Delta$ contours are not shown but the level of tuning at the boundaries of the excluded regions can be easily estimated through $y_{L4}$, which is fixed on the lines, thanks to the fact that $y_{Lt}$ is constant and approximately equal to $ y_t$ in the whole region.

As can be seen from fig.~\ref{fig:4plet-14+1_8TeV}, the run 1 LHC searches completely exclude partner masses
$m_4 \lesssim 800\ \mathrm{GeV}$. The exclusions can reach $m_4 \simeq 1\ \mathrm{TeV}$ for sizable values
of $c_t$ (namely $|c_t| \gtrsim 3$). These bounds are yet not able to exclude the natural regions of the
parameter space, indeed many configurations with a minimal amount of tuning $\Delta \sim 10$ are still viable.

\begin{figure}
\centering
\includegraphics[height=.31\textwidth]{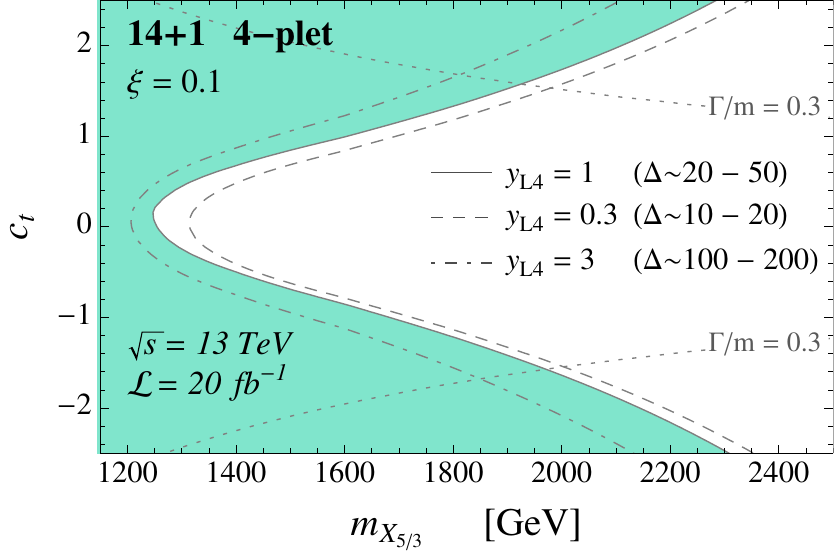}
\hspace{1em}
\includegraphics[height=.31\textwidth]{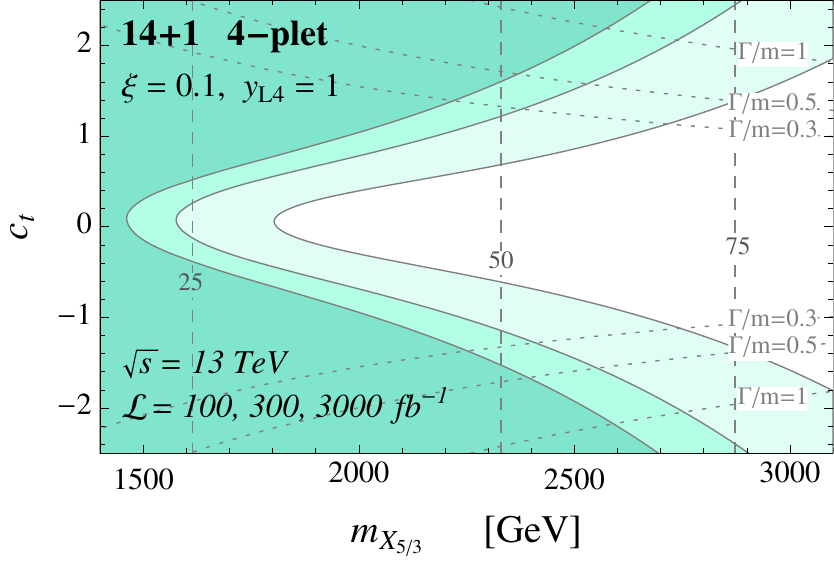}
\caption{Expected exclusions in the $14 + 1$ model with a light composite fourplet
for the $13\ \mathrm{TeV}$ LHC data in the $(m_{X_{5/3}}, c_t)$ plane for $\xi = 0.1$.
On the left panel: exclusions for ${\mathcal L} = 20\ \mathrm{fb}^{-1}$ integrated luminosity.
The green region shows the excluded points for $y_{L4} = 1$, while the dot-dashed and dashed contours correspond to
$y_{L4} = 0.3$ and $y_{L4} = 3$. The dotted contours denote the regions in which $\Gamma(X_{5/3})/m_{x_{5/3}}\geq 0.3$.
On the right panel: exclusions at the $13\ \mathrm{TeV}$ LHC for $y_{L4} = 1$ with high integrated luminosity
(${\mathcal L} = 100, 300, 3000\ \mathrm{fb}^{-1}$). The vertical dashed lines show the amount of tuning $\Delta$.
The dotted contours denote the regions with $\Gamma(X_{5/3})/m_{x_{5/3}}\geq 0.3, 0.5, 1$.
}
\label{fig:4plet-14+1_13TeV}
\end{figure}

The run 2 data will be able to probe a significantly larger part of the parameter space. Already with the
first ${\mathcal L} = 20\ \mathrm{fb}^{-1}$ of integrated luminosity, masses $m_4 \lesssim 1.2\ \mathrm{TeV}$ will be completely
covered. From the point of view of the fine-tuning, however, regions with low tuning, $\Delta \sim 1/\xi 
\sim 10$, will still
be open. A significant improvement will be obtained at the end of the LHC program, which will allow to fully
exclude resonance masses $m_4 \lesssim 1.8\ \mathrm{TeV}$ corresponding to a {\it few}$\,\%$ tuning ($\Delta \sim 30$).

Before concluding the discussion it is also interesting to notice that the width of the composite resonances
is small in the whole parameter space region accessible by the current searches. This will not be the case
any more for the run 2 LHC. In that case, in the regions with $m_4 \gtrsim 2\ \mathrm{TeV}$ and sizable
single-production couplings, the width of the $X_{5/3}$ resonance can become significant (see the dotted gray
lines in fig.~\ref{fig:4plet-14+1_13TeV})
and the narrow width approximation could not be valid any more, requiring a different search strategy.

\subsubsection*{Comparison with VLQ}

To conclude the discussion it is important to stress a difference between the composite Higgs top partners
and the VLQ's \cite{delAguila:2000aa,delAguila:2000rc,AguilarSaavedra:2009es,Aguilar-Saavedra:2013qpa}. As we saw before, the top partners couplings with the SM fermions receives significant, typically dominant, contributions from higher-order operators  and in particular from the $d_\mu$-symbol term in eq.~(\ref{eq:d_e_symbol}). VLQ's are instead described by a renormalizable effective Lagrangian and their couplings to the SM states only originate from the usual gauge interactions after the rotation to the mass-eigenstate basis. This makes that the strength of the top partners coupling, and in turn of the single production rate, is expected to be smaller for a VLQ than for a top partner, as we will show below.

For definiteness we consider a scenario with only an exotic $\SU(2)_L$ doublet with hypercharge
$Y = 7/6$, which we denote by $\Psi_{7/6} = (X_{5/3}, X_{2/3})$. The results we will derive are however valid in a generic set-up.
The effective Lagrangian describing this scenario is~\cite{AguilarSaavedra:2009es}
\begin{eqnarray}
{\mathcal L}_{VLQ} &=&  i\, \overline q_L \Dslash q_L + i\, \overline t_R \Dslash t_R + i\overline \Psi_{7/6} \Dslash \Psi_{7/6}
- M_{7/6} \overline \Psi_{7/6} \Psi_{7/6}
\nonumber\\
&& - y_t \overline q_L H^c t_R - y_{7/6} \overline \Psi_{7/6} H t_R + \mathrm{h.c.}\,,
\end{eqnarray}
where $H^c = i \sigma^2 H^*$ is the conjugate Higgs doublet. The $y_{7/6}$ parameter controls the mixing between the
SM quarks and the resonances $\Psi_{7/6}$. In particular it induces a mixing between
the $t_R$ field and the $X_{2/3\,R}$ component, whose size is controlled by the mixing angle $\phi_{\rm{VLQ}}$
\begin{equation}
\tan \phi_{\rm{VLQ}} = \frac{y_{7/6} v}{\sqrt{2} M_{7/6}}\,.
\end{equation}
At the same time the $y_{7/6}$ parameter controls the single production coupling of the $X_{5/3}$ resonance. At leading
order in the $y_{7/6}/M$ expansion, the $W X_{5/3} t_R$ coupling in the unitary gauge reads
\begin{equation}
g^{\rm{VLQ}}_{X_{5/3} t_R} = \frac{g}{2} c_R^{X_{5/3}W} = \frac{g}{2} \sqrt{2} \sin \phi_{\rm{VLQ}}\,.
\end{equation}
We can thus see that the $W X_{5/3} t_R$ coupling has an absolute upper bound in the VLQ scenarios ($c_R^{X_{5/3}W} \leq \sqrt{2}$)
and is tightly connected to the mixing between the SM states and the additional resonances.

The situation can be instead different in the composite Higgs models. In the case of the exotic $X_{5/3}$ resonance
in the $14 + 1 $ model, the leading contribution to the single production coupling (see eq.~(\ref{eq:coupl_X53}))
comes from the $d_\mu$ symbol term in eq.~(\ref{eq:fourplet_14+1}), which is a derivative interaction.
The $c_t$ coefficient can thus be sizable without generating a large mass mixing between the $t_R$ and
the composite resonances.

\begin{figure}
\centering
\includegraphics[height=.31\textwidth]{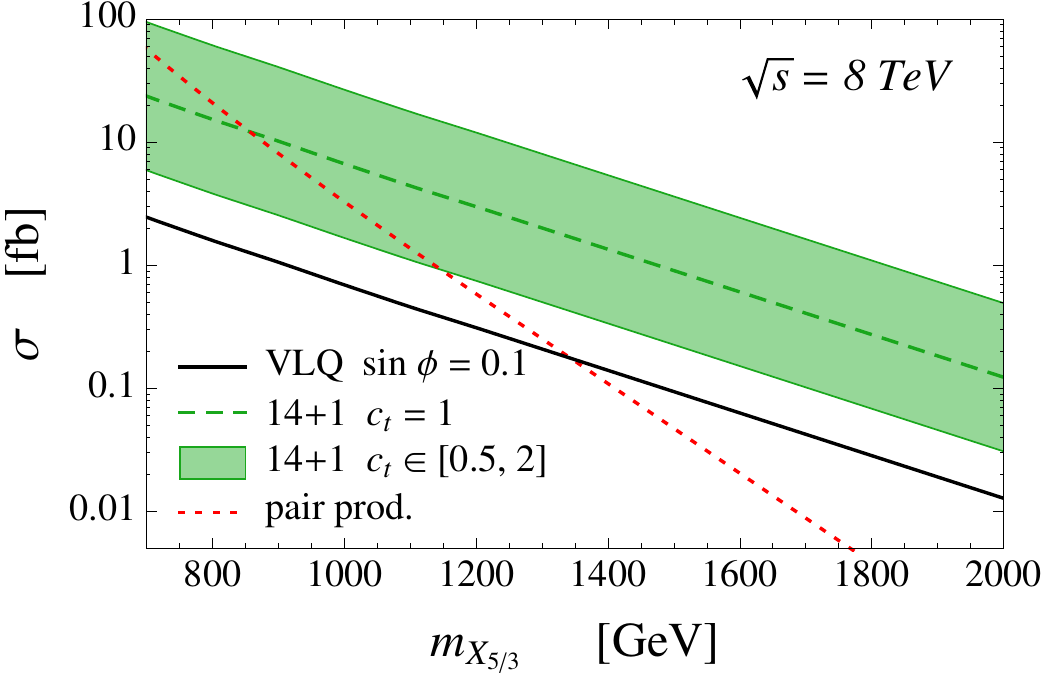}
\hspace{1em}
\includegraphics[height=.31\textwidth]{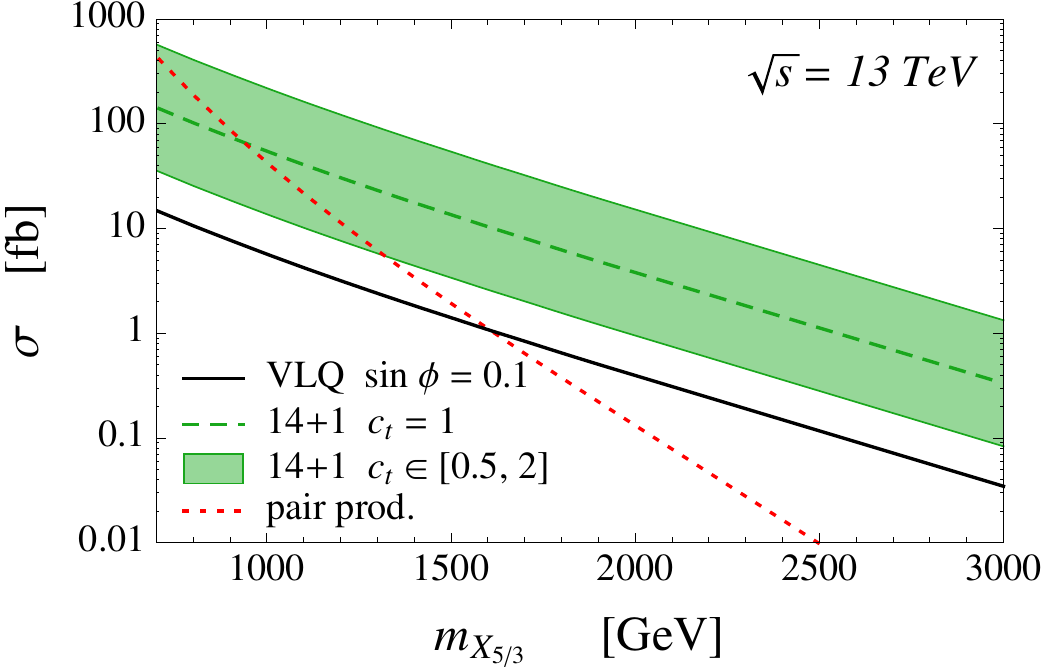}
\caption{Comparison between the production cross sections of the exotic $X_{5/3}$ state in the $14+1$ model
and in the VLQ scenarios. The left and right panels show the production cross sections at $8\ \mathrm{TeV}$
and $13\ \mathrm{TeV}$ respectively. The green band correspond to the single production channel for a top partner
with $y_{L4} = 0$ and $c_t \in [0.5, 1]$. The black line shows the single production cross section for a VLQ with
mixing angle $\sin \phi_{\textrm{VLQ}} = 0.1$. The dotted red line correspond to the universal QCD pair production cross section.
}
\label{fig:comparison_VLq}
\end{figure}

A comparison between the $X_{5/3}$ production cross sections in the $14+1$ model and in the VLQ scenario
is shown in fig.~\ref{fig:comparison_VLq}. For the VLQ case, following ref.~\cite{AguilarSaavedra:2009es}, we fixed the $t_R$
mixing angle to the value $\sin \phi_{\textrm{VLQ}} = 0.1$. For the composite Higgs case, instead,
we varied $c_t$ in the range $[0.5, 2]$ and we fixed $y_{L4} = 0$,
minimizing the mixing between the SM states and the composite resonances. Notice that the dependence on $y_{L4}$
is in any case quite limited, given that the leading single production coupling is independent of $y_{L4}$ (see eq.~(\ref{eq:coupl_X53})).
One can see from the numerical results that the single production cross section in the $14+1$ model is typically
one order of magnitude larger than the benchmark VLQ one.

\subsection{Light singlet}

We can now discuss the scenario in which the only light top partner is an $\SO(5)$ singlet $\psi_1$.
The leading operators in the effective Lagrangian can be written as
\begin{eqnarray}
{\mathcal L} &=& i \overline q_L \slashed D q_L + i \overline t_R \slashed D t_R
+ i \overline \psi_1 (\slashed D + i \slashed e) \psi_1 - m_1 \overline \psi_1 \psi_1 \nonumber\\
&& +\left(\frac{y_{Lt}}{2} f (U^t \overline q_L^{\bf 14} U)_{55} t_R + \frac{y_{L1}}{2} f (U^t \overline q_L^{\bf 14} U)_{55} \psi_1
+ \mathrm{h.c.}\right)\,.
\end{eqnarray}

In this scenario the top mass, at leading order in the $v/f$ expansion, is simply given by
\begin{equation}
m_{top}^2 = \frac{1}{2} y_{Lt}^2 f^2 \xi\,.
\end{equation}
Obviously, the spectrum of the composite states includes only a light singlet, which, following the standard notation,
we denote by $\widetilde T$. Its mass is given by
\begin{equation}
m_{\widetilde T} \simeq m_1 \left[1 + \frac{y_{L1}^2 f^2}{4 m_1^2} \xi + \cdots\right]\,.
\end{equation}

\begin{figure}
\centering
\includegraphics[height=.31\textwidth]{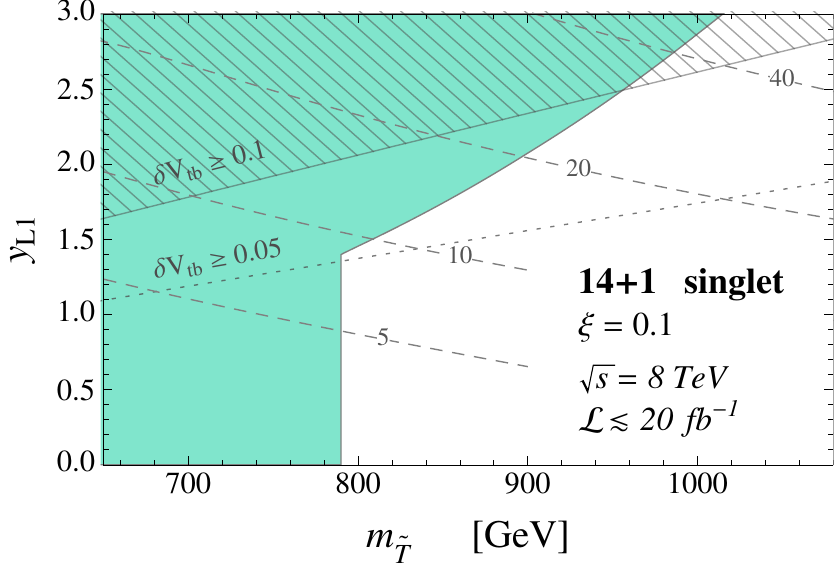}
\hspace{1em}
\includegraphics[height=.31\textwidth]{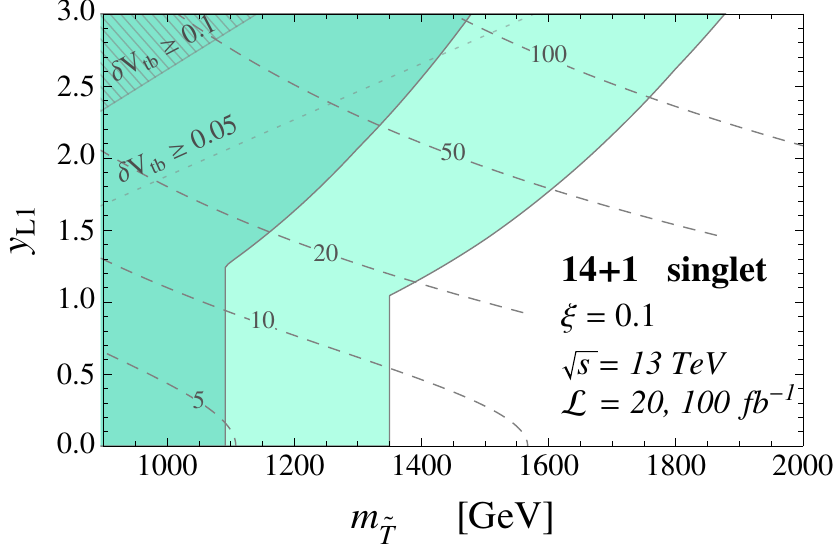}
\caption{Exclusions for the $14 + 1$ model with only a light composite singlet.
On the left panel: bounds obtained from the $8\ \mathrm{TeV}$ LHC data. On the right panel:
estimates of the exclusions for the $13\ \mathrm{TeV}$ LHC with ${\mathcal L} = 20\ \mathrm{fb}^{-1}$ (dark green) and
${\mathcal L} = 100\ \mathrm{fb}^{-1}$ (light green) integrated luminosity.
The results are presented in the $(m_{\widetilde T}, y_{R1})$ plane for $\xi = 0.1$.
The striped region corresponds to the points with $\delta V_{tb} \geq 0.1$; the corresponding bound
for $\delta V_{tb} \geq 0.05$ is shown by the dotted lines. The dashed lines show the estimate of the amount of tuning
obtained by using eq.~(\ref{eq:tuning_14+1_sing}).}
\label{fig:singlet-14+1}
\end{figure}

Let us now consider the LHC bounds. In the present set-up there is only one free parameter, $y_{L1}$,
while $y_{Lt}$ is fixed by the top mass. The exclusions from the $8\ \mathrm{TeV}$ LHC data
and an estimate of the reach for a $13\ \mathrm{TeV}$ run are shown in fig.~\ref{fig:singlet-14+1}.
In this set-up the single production coupling is strongly correlated to the value of the $y_{L1}$ mixing
\begin{eqnarray}
g_{\widetilde T b_L} &=& \frac{g}{2} c_L^{\widetilde T W} = \frac{g}{2} y_{L1} \frac{v}{m_{\widetilde T}}\,,\\
g_{\widetilde T t_L} &=& \frac{g}{2} c_L^{\widetilde T Z}
= - \frac{1}{\sqrt{2} c_w} \frac{g}{2} y_{L1} \frac{v}{m_{\widetilde T}}\,.
\end{eqnarray}
As one can see from the plots, for small $y_{L1}$ the exclusion bounds are independent of the value of the
elementary/composite mixing since they are driven by QCD pair production. The current exclusions are around
$m_{\widetilde T} \simeq 800\ \mathrm{GeV}$ and will reach $m_{\widetilde T} \simeq 1.4\ \mathrm{TeV}$
at the $13\ \mathrm{TeV}$ LHC with ${\mathcal L} = 100\ \mathrm{fb}^{-1}$ integrated luminosity.
For larger values of $y_{L1}$, $y_{L1} \gtrsim 1$, the bounds from single production instead become competitive.
In this region of the parameter space masses $m_{\widetilde T} \simeq 1\ \mathrm{TeV}$ can already be excluded
and the bounds could reach $m_{\widetilde T} \simeq 2\ \mathrm{TeV}$ at the run 2 LHC.

The amount of tuning in this scenario can be estimated as
\begin{equation}\label{eq:tuning_14+1_sing}
\Delta \simeq (y_{L1}^2 + y_{Lt}^2) \left(\frac{m_{\widetilde T}}{450\ \mathrm{GeV}}\right)^2\,.
\end{equation}
The explicit result show that regions with minimal tuning ($\Delta \sim 1/\xi \sim 10$ ) will still be allowed
at LHC run 2 with ${\mathcal L} = 100\ \mathrm{fb}^{-1}$. They will be presumably completely tested with the high-luminosity
LHC upgrade, pushing the level of tuning to the limit $\Delta \gtrsim 20$.

An interesting complementary bound on the parameter space of the light-singlet scenario comes from the measurement
on the $V_{tb}$ element of the CKM matrix. If this scenario, the single production coupling of the composite singlet
is tightly correlated to the deviations in the $\overline t_L \slashed W b_L$ coupling~\cite{Panico:2015jxa}
\begin{equation}\label{eq:gTb_Vtb}
g_{\widetilde T b_L} = g \sqrt{\delta V_{tb} - \delta V_{tb}^2/2}\,,
\end{equation}
where $\delta V_{tb} = 1 - |V_{tb}|$. This relation implies that sizable values of the single production coupling are
necessarily accompanied by large corrections in $V_{tb}$. The current experimental measurements
constrain $V_{tb}$ to the range $|V_{tb}| = 1.021 \pm 0.032$~\cite{pdg}. Taking into account the fact that in the present
set-up $|V_{tb}| \leq 1$, the experimental bound implies $g_{\widetilde T b_L} \leq 0.21\, g$ at the $2\sigma$ level.
Obviously, if additional relatively light resonances are present, the relation in
eq.~(\ref{eq:gTb_Vtb}) may be modified and larger values of $g_{\widetilde T b_L}$ could be
compatible with sufficiently small deviations in $V_{tb}$. We will discuss such a possibility in Section~\ref{sec:2site}.
This however would probably require
a certain degree of additional tuning. From the exclusion plots in fig.~\ref{fig:singlet-14+1}, it can be seen that
the constraints from the $V_{tb}$ measurements ($\delta V_{tb} \lesssim 0.05$) exclude the region in which single production
can contribute to the direct bounds. The situation will change with the run 2 LHC, for which, in the absence of
significant improvements in the $V_{tb}$ measurements, the bounds coming
from $\delta V_{tb}$ will be significantly weaker than the direct searches in single production.

\section{Double Tuning: \boldmath$\mathbf{5+5}$}\label{sec:DT}

The second class of models we consider is the one that contains the ``double tuning'' scenarios. As a representative
model we consider the $5 + 5$ set-up, in which the $q_L$ doublet and the $t_R$ singlet are realized as
elementary states mixed to  composite operators transforming in the fundamental, the $\bf 5$, representation of $\SO(5)$.
Under the unbroken $\SO(4)$ subgroup the $\bf 5$ representation decomposes as
\begin{equation}
{\bf 5} = {\bf 4} \oplus {\bf 1}\,,
\end{equation}
thus the top partners can belong to the fourplet or singlet $\SO(4)$ representation. As we did for the $14 + 1$ model,
in this section we focus on two simplified limits of the $5 + 5$ scenario in which only one multiplet of top partners
is light. An analysis of a more complete scenario including at the same time both multiplets will be presented in
sect.~\ref{sec:2site}.

\subsection{Light fourplet}

The effective Lagrangian describing the $5 + 5$ model with a light fourplet is given by
\begin{equation}\label{eq:5+5_4plet_Lagrangian}
{\mathcal L} = i \overline q_L \slashed D q_L + i \overline t_R \slashed D t_R
+ i \overline \psi_4 (\slashed D - i \slashed e) \psi_4 - m_4 \overline\psi_4 \psi_4
+ \left(y_{L4} f (\overline q_L^{\bf 5} U)_i \psi_4^i + y_{R4} f (\overline t_R^{\bf 5} U)_i \psi_4^i + \mathrm{h.c.}\right)\,.
\end{equation}
In the above equation $q_L^{\bf 5}$ and $t_R^{\bf 5}$ denote the embedding of the elementary fields in the fundamental
$\SO(5)$ representation, namely
\begin{equation}
q_L^{\bf 5} = \frac{1}{\sqrt{2}} \left(
\begin{array}{c}
-i b_L\\
- b_L\\
-i t_L\\
t_L\\
0
\end{array}
\right)\,,
\qquad \quad
t_R^{\bf 5} = \left(
\begin{array}{c}
0\\
0\\
0\\
0\\
t_R
\end{array}
\right)\,.
\end{equation}

The top mass at leading order in $v/f$ is given by
\begin{equation}\label{eq:top_mass_5+5}
m_{top}^2 = \frac{1}{2}\frac{y_{L4}^2 y_{R4}^2 f^4}
{m_4^2 + y_{L4}^2 f^2} \xi\,,
\end{equation}
while the masses of the heavy charge-$2/3$ fermionic resonances are
\begin{eqnarray}
m_{X_{2/3}} &=& m_4\left[ 1 + \frac{y_{R4}^2 f^2}{4 m_4^2} \xi + \cdots\right]\,,\label{eq:mX23_elemtR}\\
m_{T} &=& \sqrt{m_4^2 + y_{L4}^2 f^2}\left[ 1 - \frac{(y_{L4}^2 - y_{R4}^2) f^2}
{4 m_4^2} \xi + \cdots\right]\,.\label{eq:mT_elemtR}
\end{eqnarray}
Let us now consider the charge $-1/3$ states. In our model we did not include a right-handed
bottom component, therefore the $b_L$ remains in the spectrum as a massless field.
In addition to the $b_L$, the model contains also a heavy $B$ whose mass is given by
\begin{equation}\label{eq:mB_elemtR}
m_B = \sqrt{m_4^2 + y_{L4}^2 f^2}\,.
\end{equation}
Finally the exotic $X_{5/3}$ state does not mix with any other resonance and has a mass $m_{X_{5/3}} = m_4$,
which does not receive any shift after EWSB.

\begin{figure}
\centering
\includegraphics[width=0.3\textwidth]{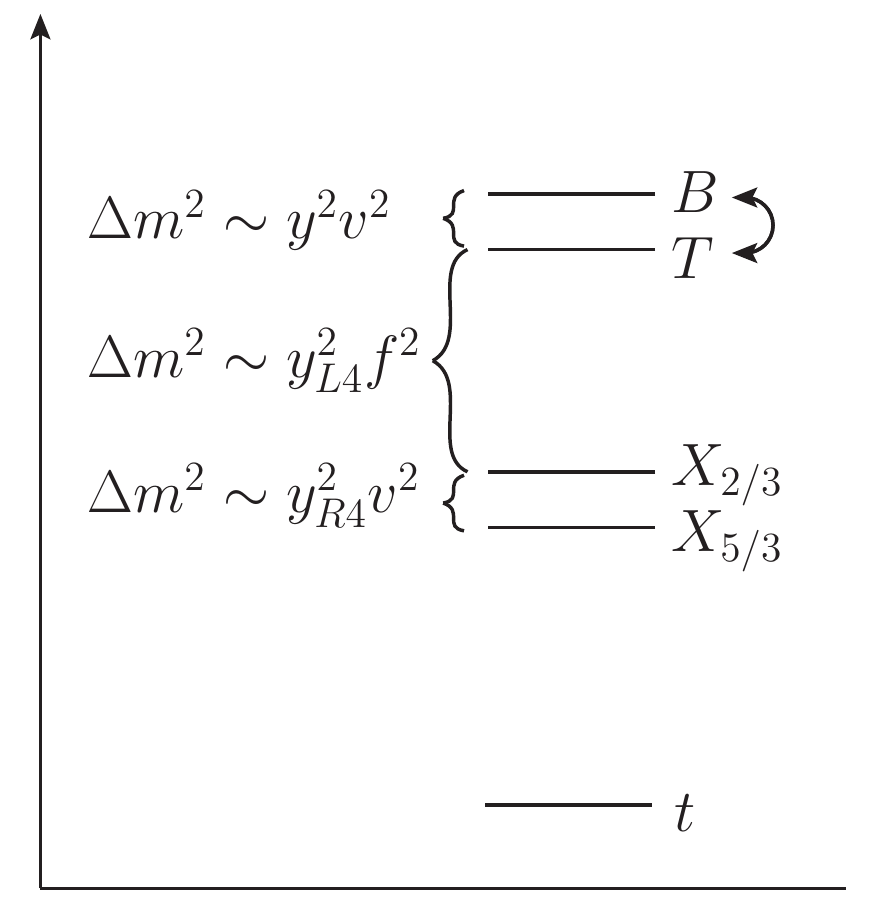}
\caption{Typical mass spectrum of the quadruplet states in the $5+5$ model.}
\label{fig:mass_spectrum_elemtR}
\end{figure}

The spectrum of the composite resonances resembles quite closely the one we found in the $14 + 1$ model
(see Fig.~\ref{fig:mass_spectrum_comptR}). It consists in two approximate $\SU(2)_L$ doublets,
the $(T, B)$ and the $(X_{5/3}, X_{2/3})$, separated by a mass splitting of order $\Delta m^2 \sim y_{L4}^2 f^2$.
The splitting inside each doublet is instead much smaller, of order $\Delta m^2 \sim y^2 v^2$, where
$y$ collectively denotes the elementary/composite mixings.
The only difference with respect to the $14 + 1$ model is the fact that the two states in the $(X_{5/3}, X_{2/3})$ doublet
are not fully degenerate, but instead are split by EWSB effects. The structure of the mass spectrum is schematically
shown in fig.~\ref{fig:mass_spectrum_elemtR}.

\begin{figure}
\centering
\includegraphics[height=.3\textwidth]{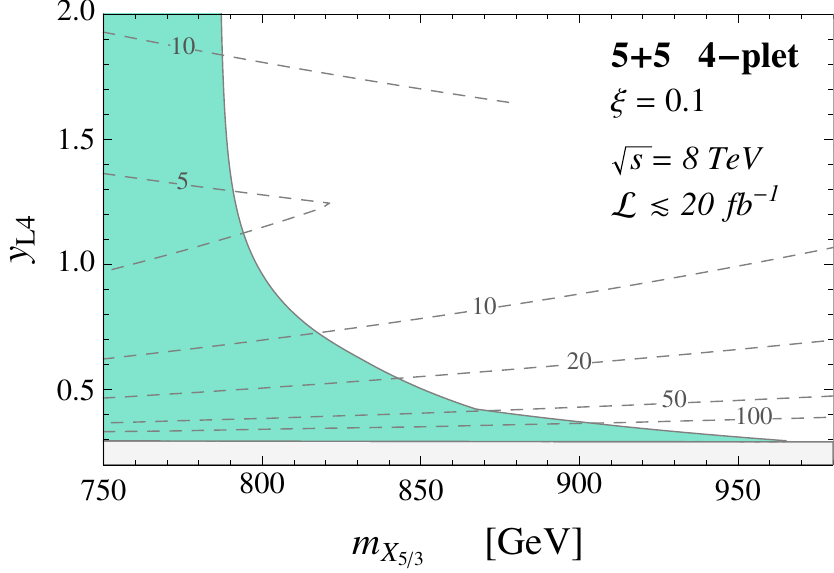}
\caption{Exclusion plot for the $5 + 5$ model with only a light composite 4-plet for the $8\ \mathrm{TeV}$
LHC data in the $(m_{X_{5/3}}, y_{L4})$ plane. The green region shows the excluded points for $\xi = 0.1$.
The shaded gray area is not theoretically allowed since the correct top mass can not be reproduced.
The dashed contours show the amount of tuning $\Delta$ estimated by using eq.~(\ref{eq:tuning_5+5_4plet}).}
\label{fig:4plet-5+5_8TeV}
\end{figure}

The Lagrangian in eq.~(\ref{eq:5+5_4plet_Lagrangian}) has only three free parameters, namely the elementary/composite mixings,
$y_{L4}$ and $y_{R4}$, and the mass of the 4-plet, $m_4$. By requiring the correct top mass to be reproduced
we can fix one of the parameters, which we conveniently chose to be the right-handed mixing $y_{R4}$.
The experimental constraints can thus be presented as exclusions in the $(m_{X_{5/3}}, y_{L4})$ plane.
The current bounds from the $8\ \mathrm{TeV}$ LHC data are shown in fig.~\ref{fig:4plet-5+5_8TeV}.
The constraints are weaker for large values of the left-handed elementary/composite mixing ($y_{L4} \gtrsim 1$),
where masses $m_{X_{5/3}}$ below $800\ \mathrm{GeV}$ are fully excluded. At smaller values of $y_{L4}$ the
exclusions become stronger and can reach up to $m_{X_{5/3}} \simeq 950\ \mathrm{GeV}$.
The increase in the bounds comes from two simultaneous effects: the contributions of the $B$ state and
the single production of the $X_{5/3}$ resonance. The latter effect is the main one and determines almost completely
the enhancement in the bounds. The relevant single-production coupling is approximately given by
\begin{equation}
g_{X_{5/3}t_R} = \frac{g}{2} c_R^{X_{5/3} W}
\simeq -\frac{g}{2} \frac{v}{f} \frac{f}{m_{X_{5/3}}} y_{R4} \simeq -\frac{g}{2} \frac{v}{f} \frac{f}{m_{X_{5/3}}}
y_{top} \frac{\sqrt{m_{X_{5/3}}^2/f^2 + y_{L4}^2}}{y_{L4}}\,.
\end{equation}
The dependence on $y_{L4}$ in the last equality explains why single production is more relevant for
small left-handed elementary/composite mixing.
The effects due to the $B$ resonance is analogous to what we discussed in the $14 + 1$ model with a light 4-plet.
At small values of $y_{L4}$ the mass split between the $X_{5/3}$ and the $B$ decreases (see eq.~(\ref{eq:mB_elemtR})),
thus the production cross section of the two resonances becomes comparable.

\begin{figure}
\centering
\includegraphics[height=.3\textwidth]{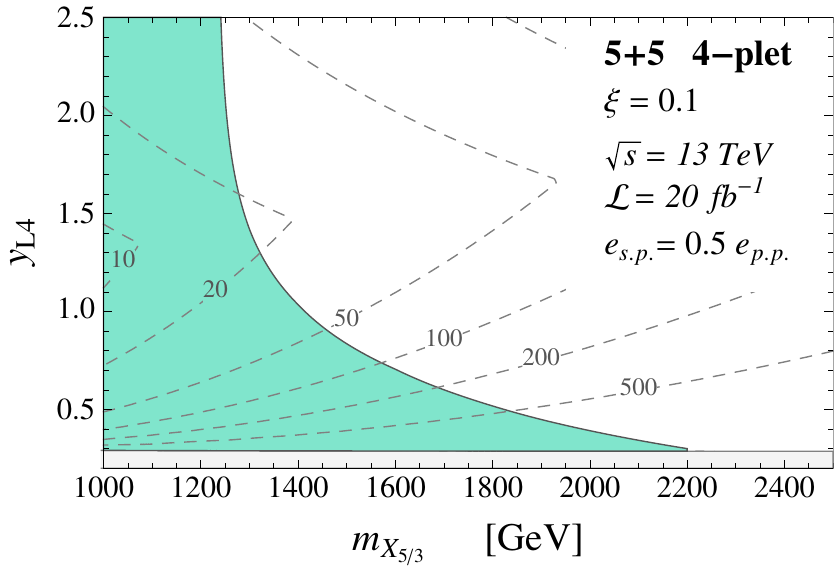}
\hspace{1em}
\includegraphics[height=.3\textwidth]{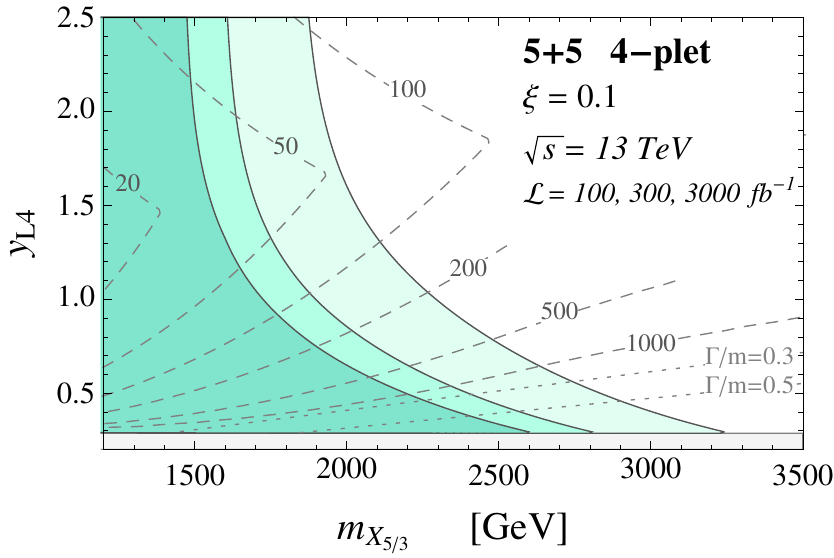}
\caption{Estimated exclusion on the $5 + 5$ model with only a composite 4-plet
for the $13\ \mathrm{TeV}$ LHC run. The results are shown in the $(m_{X_{5/3}}, y_{L4})$ plane for the choice $\xi = 0.1$.
On the left panel, the green area shows the excluded region for ${\mathcal L} = 20\ \mathrm{fb}^{-1}$ integrated luminosity
under the assumption that the signal efficiency of single production processes is $50\%$ of to the pair production one,
$e_{s.p.} = 0.5\, e_{p.p.}$.
On the right panel the green areas show the expected exclusions
for ${\mathcal L} = 100, 300, 3000\ \mathrm{fb}^{-1}$ integrated luminosity (assuming $e_{s.p.} = 0.5\, e_{p.p.}$).
The dotted contours denote the regions with $\Gamma(X_{5/3})/m_{X_{5/3}} = 0.3, 0.5$.
On both panels the dashed contours show the amount of tuning $\Delta$ estimated by using eq.~(\ref{eq:tuning_5+5_4plet}).
}
\label{fig:4plet-5+5_13TeV}
\end{figure}

Similar effects are present in the estimates for the exclusions at the $13\ \mathrm{TeV}$ LHC.
In this case, as shown in fig.~\ref{fig:4plet-5+5_13TeV}, the high-luminosity LHC program should be able to probe masses
up to $m_{X_{5/3}} \sim 2 - 3\ \mathrm{TeV}$.

The tuning estimate in the $5 + 5$ model follows a slightly different pattern than in the $14 + 1$ case.
In the present set-up, indeed, the Higgs potential receives contributions from the left-handed and the right-handed
elementary/composite mixings, since both mixings break the Goldstone symmetry.
The amount of tuning can thus be estimated as~\footnote{Notice that in the tuning estimate we did not sum the
contributions from the left-handed and right-handed mixings. This slightly more conservative choice is motivated
by the fact that in explicit models the dependence on the mixings factorizes at leading order and the cancellation
comes form a tuning between the values of the two parameters~\cite{Panico:2011pw,Matsedonskyi:2012ym}.}
\begin{equation}\label{eq:tuning_5+5_4plet}
\Delta \simeq \max(y_{L4}^2, y_{R4}^2) \left(\frac{m_{X_{5/3}}}{450\ \mathrm{GeV}}\right)^2\,.
\end{equation}

The results in fig.~\ref{fig:4plet-5+5_8TeV} show that configurations with minimal amount of tuning $\Delta \sim 1/\xi \sim 10$
are still compatible with the $8\ \mathrm{TeV}$ LHC data. The high-luminosity LHC program, on the other hand,
could be able to fully exclude the parameter space region with $\Delta \lesssim 50$.

\subsection{Light singlet}

As a second scenario in the class of ``double tuning'' models, we consider the $5 + 5$ set-up with only a light singlet.
The effective Lagrangian describing this model is
\begin{equation}\label{eq:Lagr_5+5_singlet}
{\mathcal L}  =  i \overline q_L \slashed D q_L + i \overline t_R \slashed D t_R
+ i \overline \psi_1 \slashed D \psi_1 - m_1 \overline\psi_1 \psi_1
+ \left(y_{L1} f (\overline q_L^{\bf 5} U)_5 \psi_1 + y_{R1} f (\overline t_R^{\bf 5} U)_5 \psi_1 + \mathrm{h.c.}\right)\,.
\end{equation}
Analogously to the case with a light 4-plet, the effective Lagrangian contains only $3$ free parameters,
namely $y_{L1}$, $y_{R1}$ and $m_1$. One of the parameters can be fixed by requiring the correct value for the top mass,
whose approximate expression, at leading order in $v/f$, is given by
\begin{equation}
m_{top}^2 = \frac{1}{2} \frac{y_{R1}^2 y_{L1}^2 f^4}{m_1^2 + y_{R1}^2 f^2} \xi\,.
\end{equation}
The mass of the composite resonance $\widetilde T$ is instead given by
\begin{equation}
m_{\widetilde T} = \sqrt{m_1^2 + y_{R1}^2 f^2}\left[1 + \frac{(y_{L1}^2 - 2 y_{R1}^2) f^2}{4 m_1^2} \xi + \cdots\right]\,.\label{eq:mTtilde_elemtR}
\end{equation}

\begin{figure}
\centering
\includegraphics[height=.3\textwidth]{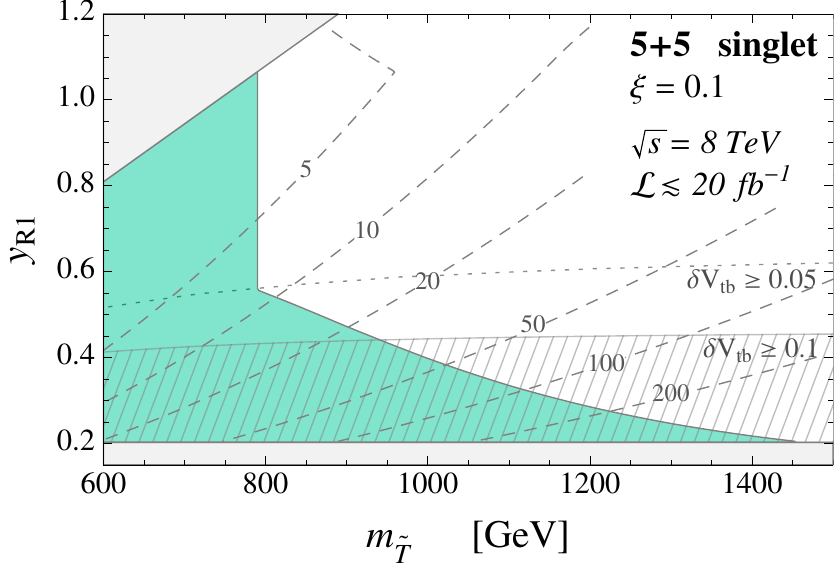}
\hspace{1em}
\includegraphics[height=.3\textwidth]{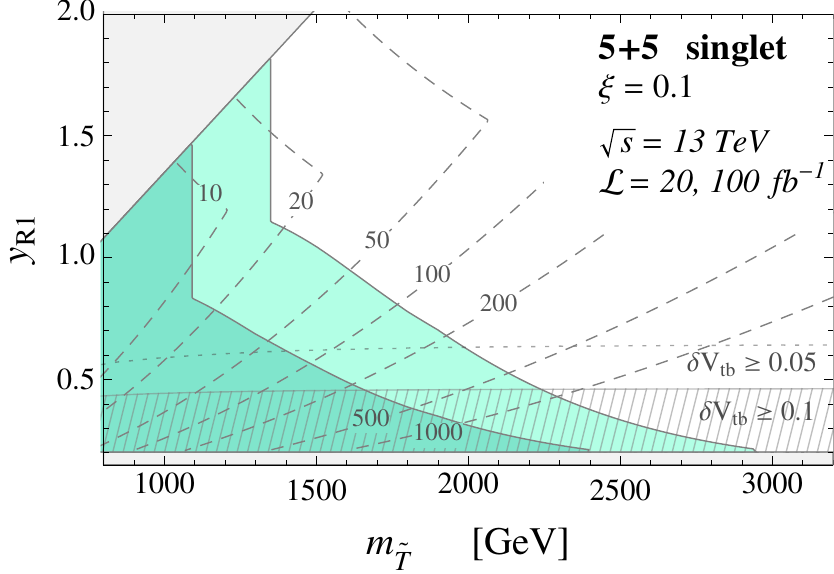}
\caption{Exclusion plot for the $5 + 5$ model with only a composite singlet with $\xi = 0.1$
The left panel shows the bounds for the $8\ \mathrm{TeV}$ LHC data, while the right panel shows
an estimate of the constraints from the $13\ \mathrm{TeV}$ LHC run with ${\mathcal L} = 20, 100\ \mathrm{fb}^{-1}$ integrated luminosity.
The green region shows the excluded points in the $(m_{\widetilde T}, y_{R1})$ plane.
The shaded gray area is not theoretically allowed. The striped region corresponds to the points
with $\delta V_{tb} \geq 0.1$, while the corresponding bound for $\delta V_{tb} \geq 0.05$
is denoted by the dotted gray line.
The dashed contours show the amount of tuning $\Delta$ estimated by using eq.~(\ref{eq:tuning_5+5_singlet}).}
\label{fig:singlet-5+5}
\end{figure}

The current bounds from the $8\ \mathrm{TeV}$ LHC data and an estimate of the exclusions in the $13\ \mathrm{TeV}$
run are shown in fig.~\ref{fig:singlet-5+5}. Pair production leads to the strongest bounds for large values of $y_{R1}$
($y_{R1} \gtrsim 0.6 - 1$). Single production becomes competitive at smaller values of the right-handed elementary/composite
mixing due to the enhancement of the $W \widetilde T_R b_L$ coupling:
\begin{equation}
g_{\widetilde T b_L} = \frac{g}{2} c_L^{\widetilde T W}
\simeq \frac{g}{2}\frac{v}{f}\frac{f y_{L1} m_1}{m_1^2 + y_{R1}^2 f^2}
\simeq \frac{g}{2} \frac{v}{f} \frac{m_1}{\sqrt{m_1^2 + y_{R1}^2 f^2}} \frac{y_{top}}{y_{R1}}\,.
\end{equation}

Analogously to the case with a light 4-plet, we can estimate the amount of tuning by the formula
\begin{equation}\label{eq:tuning_5+5_singlet}
\Delta \simeq \max\left(y_{R1}^2, y_{L1}^2\right) \left(\frac{m_{\widetilde T}}{450\ \mathrm{GeV}^2}\right)^2\,.
\end{equation}
As in the other simplified models we considered, configurations with small tuning $\Delta \simeq 10$ are still compatible with the
present LHC data. The run 2 LHC with integrated luminosity ${\mathcal L} \simeq 100\ \mathrm{fb}^{-1}$ could
completely cover the parameter space region with $\Delta \lesssim 20$.

In addition to the direct exclusions coming from the LHC searches, complementary bounds on the parameter space
can be derived from the measurement of the $V_{tb}$ matrix element. In complete analogy to what we discussed in the
$14 + 1$ model, the $\widetilde T$ single production coupling and the deviations
in the $V_{tb}$ matrix element are related by eq.~(\ref{eq:gTb_Vtb}). From the left panel in fig.~\ref{fig:singlet-5+5},
one can see that, in the case of the $8\ \mathrm{TeV}$ LHC data, the parameter space region where single production leads
to a significant bound ($y_{R1} \lesssim 0.6$)
is already excluded by the current bounds $\delta V_{tb} \lesssim 0.05$. In the run 2 LHC, on the other hand, single production
could probe some regions of the parameter space which are not covered by the present $V_{tb}$ measurements (right panel
of fig.~\ref{fig:singlet-5+5}).

\section{The two-site model}\label{sec:2site}

As a last scenario we consider the $2$-site construction proposed in
refs.~\cite{Panico:2011pw,Matsedonskyi:2012ym}.\footnote{For analogous constructions see also ref.~\cite{DeCurtis:2011yx}.}
This set-up includes an extended set of global symmetries that stabilize the Higgs potential through a collective
breaking mechanism. In the following for definiteness we will focus on a $2$-site realization in which the $q_L$
doublet and the $t_R$ singlet are realized as elementary states and are coupled to resonances in the fundamental
$\SO(5)$ representation. We will call this scenario the ``$5 + 5$ $2$-site model''.
This set-up is analogous to a ``deconstructed'' version of the MCHM$_5$ scenario~\cite{Contino:2006qr}.

As shown in ref.~\cite{Matsedonskyi:2012ym} (see also ref.~\cite{Panico:2015jxa} for a more detailed discussion),
the collective breaking structure ensures a partial calculability for the Higgs potential. In particular this allows,
once we fix the value of $\xi$, to extract the value of the Higgs mass as a function
of the parameters of the model (namely the resonance masses and the elementary/composite mixings).

The Lagrangian for the $5 + 5$ $2$-site model can be mapped onto the $5 + 5$ effective models
described in Section~\ref{sec:DT} and contains one layer of composite
resonances that transform as a $4$-plet and a singlet of $\SO(4)$. The complete Lagrangian includes
the terms in eqs.~(\ref{eq:5+5_4plet_Lagrangian}) and (\ref{eq:Lagr_5+5_singlet}), namely
\begin{eqnarray}
{\mathcal L} &=& i \overline q_L \slashed D q_L + i \overline t_R \slashed D t_R + i \overline \psi_4 (\slashed D - i \slashed e) \psi_4+ i \overline \psi_1 \slashed D \psi_1 - m_4 \overline \psi_4 \psi_4 - m_1 \overline \psi_1 \psi_1\nonumber\\
&& + y_L f \overline q_L^{\bf 5} U \Psi + y_R f \overline t_R^{\bf 5} U \Psi + \mathrm{h.c.}\,,\label{eq:Lagr_2-site}
\end{eqnarray}
together with an additional interaction involving the composite partners \footnote{This additional term
was not included in the original constructions of refs.~\cite{Panico:2011pw,Matsedonskyi:2012ym}.
This choice was guided by minimality and by an analogy with extra-dimensional realizations of the composite Higgs idea.
The term, however, is allowed by the symmetry structure and, as we will see in the following,
can have some phenomenological impact, so we include it in the present study.}
\begin{equation}\label{eq:Lagr_2site_dsymbol}
{\mathcal L}_{comp} = - i c \overline \psi_4^i \gamma^\mu d_\mu^i \psi_1 + \mathrm{h.c.}\,.
\end{equation}
In eq.~(\ref{eq:Lagr_2-site}) we denoted by $\Psi$ the $\SO(5)$ $5$-plet built from the $\psi_4$ and $\psi_1$ fields.
Notice that, as required by the collective structure assumption, the elementary/composite mixings involving the
$\psi_4$ and $\psi_1$ resonances are not independent parameters as in eqs.~(\ref{eq:5+5_4plet_Lagrangian})
and (\ref{eq:Lagr_5+5_singlet}), but instead they are related, $y_{L4} = y_{L1} \equiv y_L$ and $y_{R4} = y_{R1} \equiv y_R$.

The number of free parameters can be reduced by fixing the top and the Higgs mass. An approximate expression for the
top mass is given by
\begin{equation}
m_{top}^2 = \frac{1}{2} \frac{y_L^2 y_R^2 f^4 (m_4 - m_1)^2}{(m_4^2 + y_L^2 f^2)(m_1^2 + y_R^2 f^2)}\xi\,.
\end{equation}
The Higgs mass, on the other hand,  as shown in ref.~\cite{Matsedonskyi:2012ym}
(see also refs.~\cite{Marzocca:2012zn,Pomarol:2012qf}),
is simply related to the masses of the singlet ($m_{\widetilde T}$) and of the $T$ resonance inside the $4$-plet ($m_T$):
\begin{equation}\label{eq:mhmt}
m_h \simeq m_{top} \frac{\sqrt{2 N_c}}{\pi} \frac{m_T m_{\widetilde T}}{f}
\sqrt{\frac{\log(m_T/m_{\widetilde T})}{m_T^2 - m_{\widetilde T}^2}}\,,
\end{equation}
where $N_c = 3$ is the number of QCD colors.
This relation is valid with good accuracy even if the effects of other layers of resonances are taken into account,
the typical corrections being of order $20 - 30\%$. To take these effects into account 
we assume that eq.~(\ref{eq:mhmt}) is verified with $20\%$ accuracy and we associate to each point in our exclusion plots
the ``ensemble'' of configurations compatible with the $20\%$ uncertainty. We consider one point excluded only if all
the configurations in the corresponding ensemble are excluded. Using the two constraints mentioned above, we are
left with three free parameters, which can be conveniently identified with the mass of the exotic resonance $X_{5/3}$
(that coincides with the $m_4$ parameter), the left mixing angle $\phi_L$ (related to the left mixing by
$\tan \phi_L = y_L f/m_4$) and the coefficient of the $d_\mu$-symbol interaction, $c$. For each pair $(m_{X_{5/3}}, \phi_L)$
which allows to get the correct top and Higgs mass, two solutions for $m_1$ and $y_R$ are found. In order to
represent the whole parameter space on two dimensional plots, we assign the two solutions to two
distinct sets, which we denote by ``Region I'' and ``Region II''.

Before analyzing the LHC bounds, it is interesting to discuss two preliminary aspects, namely the estimate of the tuning
and the connection between the single-production couplings and the deviations in $V_{tb}$.

As we briefly mentioned before, the collective symmetry breaking structure of the $2$-site models ensures that the
potential does not develop a quadratic divergence as would be the case in a generic CCWZ construction. This fact tells us that
the top partners included in the $2$-site description are the ones responsible for cancelling the quadratic divergence and thus
they fix the size of the leading contributions to the Higgs mass. We can use this information to find a more reliable formula
for the amount of tuning
\begin{equation}\label{eq:tuning_2site}
\Delta \simeq \max(y_L^2, y_R^2) \left(\frac{M}{450\ \mathrm{GeV}}\right)^2\,,
\end{equation}
where we denoted by $M$ the largest composite mass parameter, namely $M = \max(|m_4|, |m_1|)$. The choice
of taking the maximum between the two elementary/composite mixings is now fully justified since the leading term
in the potential has the structure~\cite{Matsedonskyi:2012ym,Panico:2015jxa}
\begin{equation}
V_{y^2} \sim \frac{N_c}{16 \pi^2} M^2 f^2 \left(\frac{y_L^2}{2} - y_R^2\right) \xi\,,
\end{equation}
so that the cancellation mainly takes place by balancing the left-handed and right-handed contributions.
The choice of using the maximum between the two composite mass parameters is instead justified by the fact that
the whole set of resonances $\psi_4$ and $\psi_1$ is responsible for cancelling the quadratic divergence, which is still
present if only one $\SO(4)$ multiplet is light.

\begin{figure}
\centering
\includegraphics[width=.375\textwidth]{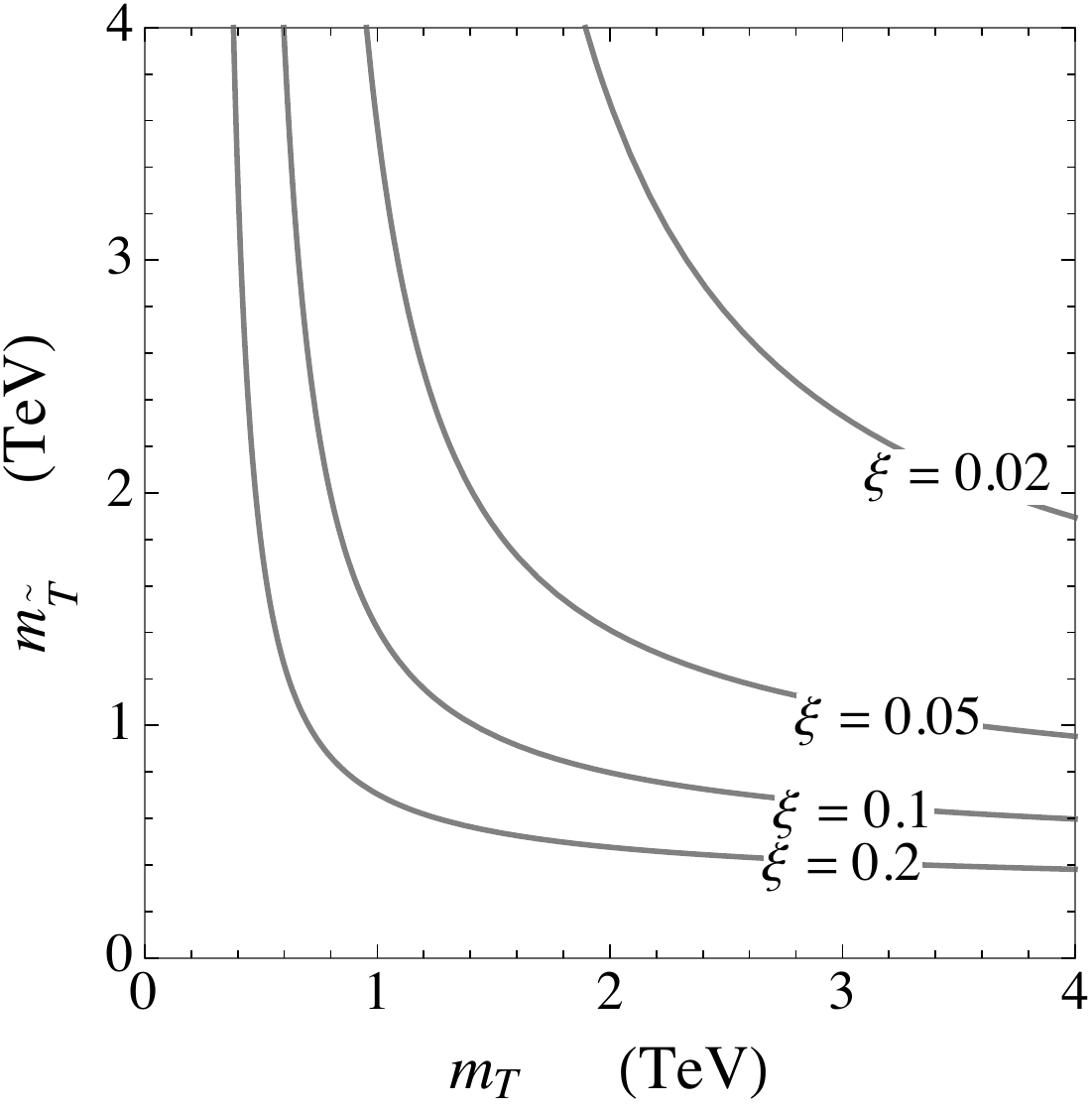}
\caption{Relation between the $T$ and $\widetilde T$ resonances masses in the $5 + 5$ $2$-site model
for different values of $\xi$. The curves are obtained by using eq.~(\ref{eq:mhmt}).}
\label{fig:mh_mt}
\end{figure}
Looking at the relation between the Higgs mass and the masses of the composite resonances in eq.~(\ref{eq:mhmt})
we can get a further insight on the amount of tuning. In order to reproduce the correct Higgs mass, the
$T$ and $\widetilde T$ masses must lie on some approximately hyperbolic curves as shown in fig.~\ref{fig:mh_mt}.
This means that the overall mass scale of the resonances, and thus the amount of tuning, is minimized
when both the $4$-plet and the singlet have similar masses, namely
\begin{equation}
m_T \sim m_{\widetilde T} \sim \frac{\pi}{\sqrt{3}} \frac{m_h}{m_{top}} f
\simeq \frac{350\ \mathrm{GeV}}{\sqrt{\xi}}\,.
\end{equation}
This expectation is confirmed by the numerical results as we will see in the following.

A second aspect that is worth discussing is the relation between the single production coupling of the $\widetilde T$
singlet and the deviations in the $V_{tb}$ CKM matrix element. We already saw that in the simplified scenarios with only
a light singlet a tight relation exists between these two quantities (see eq.~(\ref{eq:gTb_Vtb})).
This relation is a consequence of the fact that
in those set-ups the $W \widetilde T b_L$ vertex originated exclusively from the $W t_L b_L$ vertex after the rotation
to the mass eigenstate basis~\cite{Panico:2015jxa}. The situation is slightly different in the $5 + 5$ $2$-site model.
In this case an additional contribution to the $W \widetilde T b_L$ vertex comes from the $d_\mu$-symbol
interaction in eq.~(\ref{eq:Lagr_2site_dsymbol}) since the $\psi_4$ multiplet is mixed with the $b_L$ field.
A further effect comes from the mixing of the $b_L$ field with the $\psi_4$ multiplet, which determines a correction to $V_{tb}$.
The relation in eq.~(\ref{eq:gTb_Vtb}) is replaced by the following formula valid at leading order in the $v/f$ expansion
\begin{equation}
g_{\widetilde T b_L}^2 = \frac{g^2}{2} (1 - |V_{tb}|^2) + \frac{2 c^2 -1}{4} \xi\, g^2 \sin^2 \phi_L\,.
\end{equation}
We checked that this relation is in very good agreement with the numerical results.

We can now discuss the present and expected future bounds on the $2$-site model coming from the LHC searches.
Since this model contains resonances in the $4$-plet and singlet representation of $\SO(4)$, the parameter space
can be constrained by using both the searches for the exotic $X_{5/3}$ state and the ones for charge-$2/3$ resonances.
As we explained before, due to the relation between the Higgs mass and the mass of the
top partners in eq.~(\ref{eq:mhmt}), if one $\SO(4)$ multiplet is heavy, the other must necessarily be light and is the one
which determines the exclusions in this part of the parameter space. When the $4$-plet is light
the $X_{5/3}$ is always among the lightest states. Due to level repulsion effects,
if the singlet $\widetilde T$ is relatively close in mass to the $4$-plet, the lightest state can be the $X_{2/3}$ and not the $X_{5/3}$. However, even in these regions of the parameter space the strongest bounds usually come from the $X_{5/3}$
searches.

\begin{figure}
\centering
\includegraphics[width=.42\textwidth]{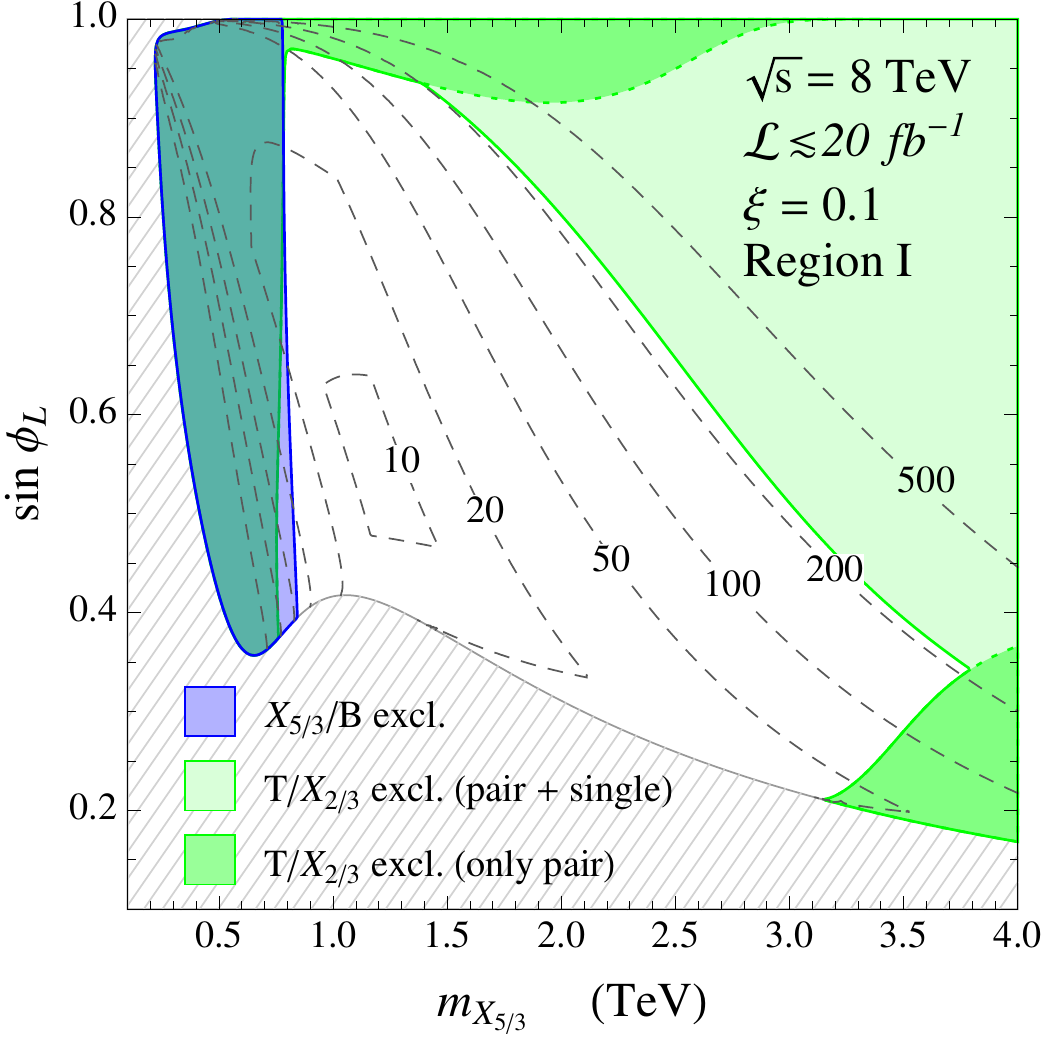}
\hspace{2em}
\includegraphics[width=.42\textwidth]{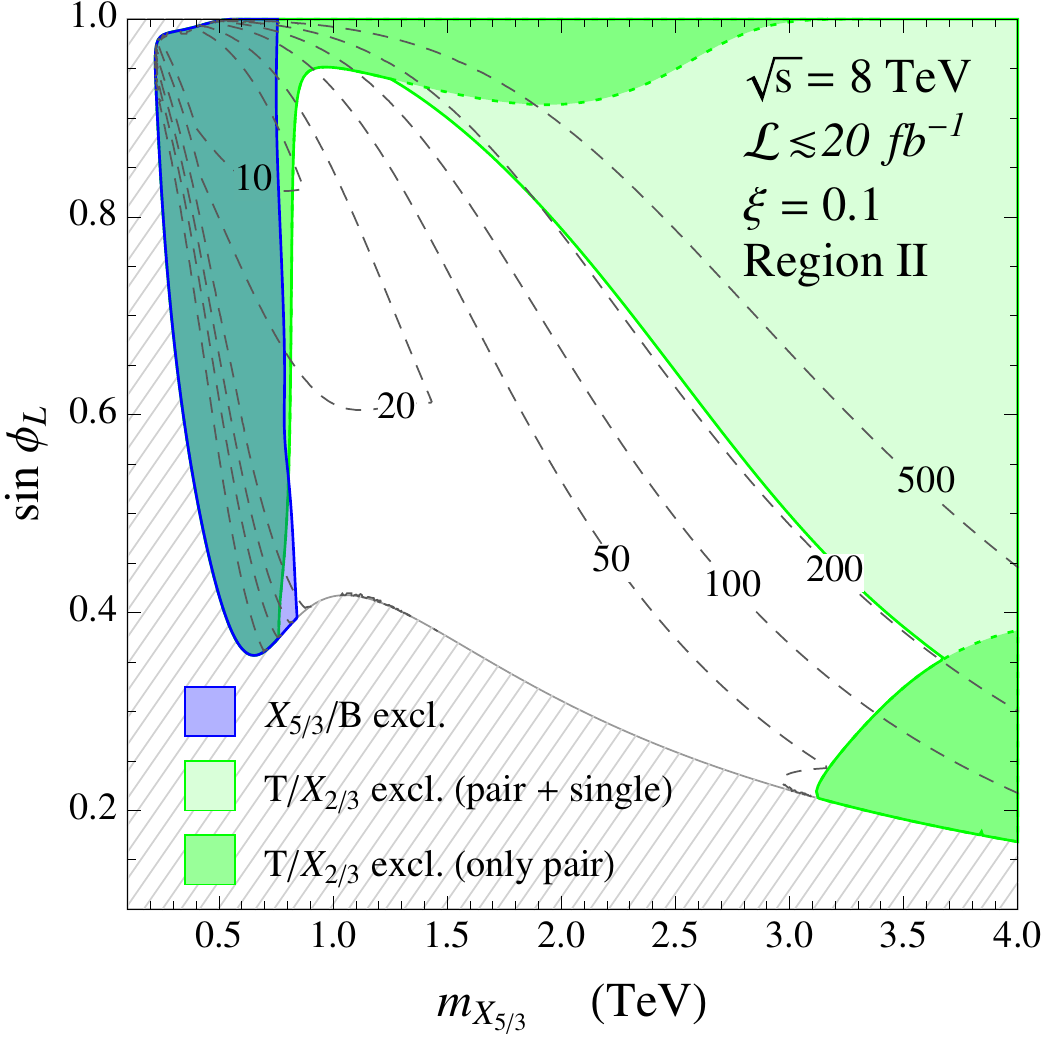}
\caption{Exclusion bounds in the $2$-site model with $\xi = 0.1$ and $c=0$ for the 8 TeV LHC data.
The left (right) panel corresponds to the Region I (Region II) of the parameter space.
The blue and green region are excluded by the searches for the exotic $X_{5/3}$
and the charge-$2/3$ resonances respectively.
The darker green region shows the exclusions on the charge-$2/3$ states if only pair production is taken
into account, while the estimates of additional constraints from single production are shown by the light
green area. The dashed contours show the amount of tuning $\Delta$ estimated by using eq.~(\ref{eq:tuning_2site}).
}
\label{fig:2-site_xi01_8TeV}
\end{figure}

The present exclusions for $\xi = 0.1$ are shown in fig.~\ref{fig:2-site_xi01_8TeV}
in the plane $(m_{X_{5/3}}, \sin \phi_L)$ for the choice $c = 0$. The current LHC data can already
exclude a non-negligible part of the parameter space, although configurations with minimal amount of tuning are still
allowed. It is interesting to notice that, if we only rely on pair production, the bounds become quite
mild, basically disappearing in the regions with a light singlet and a heavy $4$-plet ($m_{X_{5/3}} \gtrsim 1\ \mathrm{TeV}$).
The drastic change in the bounds coming from the inclusion of single production can be understood as follows.
In the configurations with a light singlet, the mass of the $\widetilde T$ resonance depends only mildly on the $m_{X_{5/3}}$
parameter (see fig.~\ref{fig:mh_mt}) and is slightly above the current pair-production bound. The mild increase in the
bound coming from single production searches (of order $200\ \mathrm{GeV}$) is thus enough to exclude all these configurations.

\begin{figure}
\centering
\includegraphics[width=.42\textwidth]{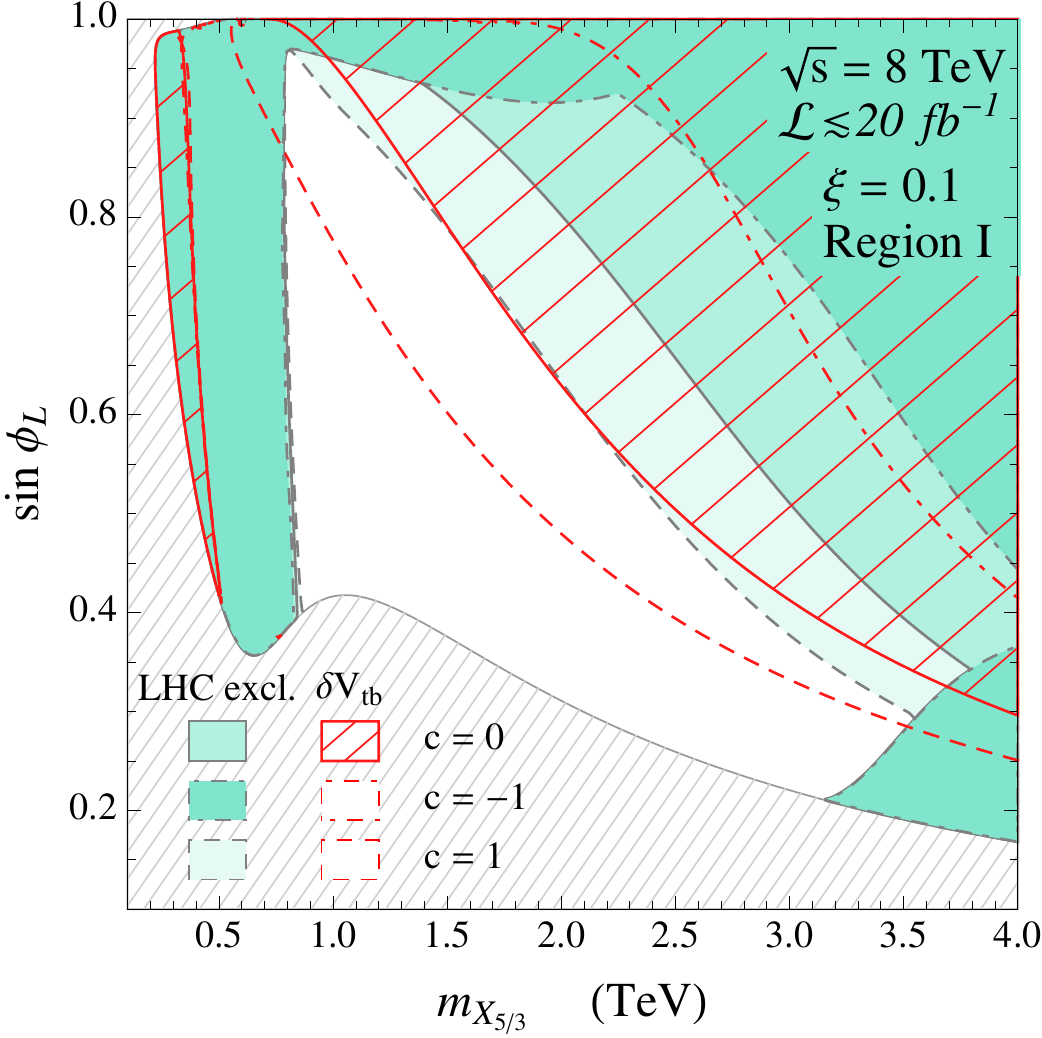}
\hspace{2em}
\includegraphics[width=.42\textwidth]{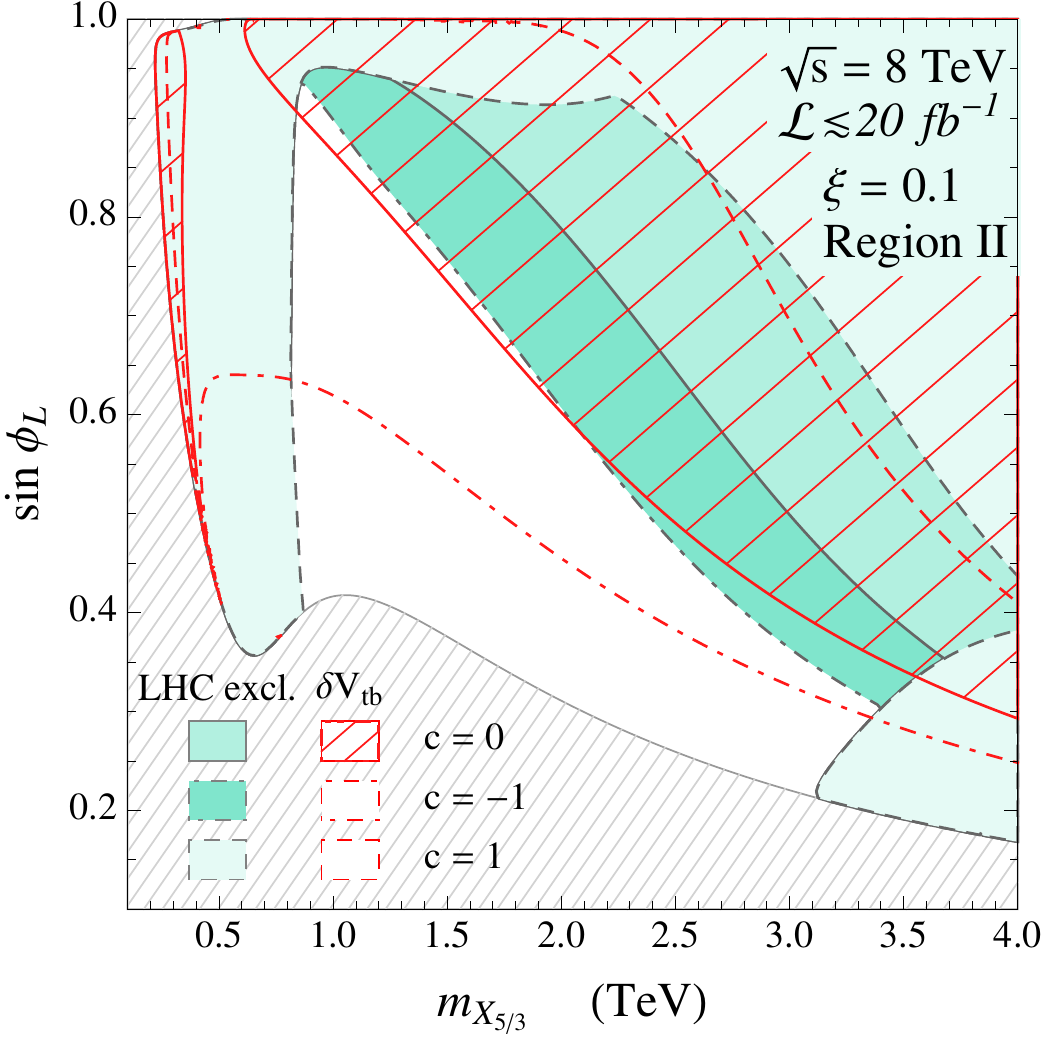}
\caption{Exclusion bounds in the $2$-site model with $\xi = 0.1$ coming from the 8 TeV LHC data
for different values of $c = 0,1, -1$.
The green regions show the bounds coming from the direct searches,
while the red contours show the $95\%$ CL constraints coming from the $V_{tb}$ measurement.}
\label{fig:2-site_xi01_8TeV_cVtb}
\end{figure}

It must be stressed that the single-production bound strongly depends on the $W \widetilde T b_L$ coupling.
As a consequence, it is sensitive to the value of the $c$ parameter. The change in the
bounds for different values of $c$, namely $c = 0, 1, -1$, is shown in fig.~\ref{fig:2-site_xi01_8TeV_cVtb}.
From the explicit results one can see that the impact of an order one variation in $c$ can significantly affect the
exclusion bounds. It must be noticed, however, that the direct bounds coming from single (and pair) production in the
configurations with a light singlet are currently barely competitive with the indirect ones coming from the measurement
of the $V_{tb}$ CKM element. The $95\%$ exclusion contours, corresponding to $\delta V_{tb} = 0.043$ are shown
by the red contours in fig.~\ref{fig:2-site_xi01_8TeV_cVtb}.
At the next LHC runs, on the other hand, the improvement in the direct searches will make the corresponding exclusions
stronger than the indirect ones.

\begin{figure}
\centering
\includegraphics[width=.42\textwidth]{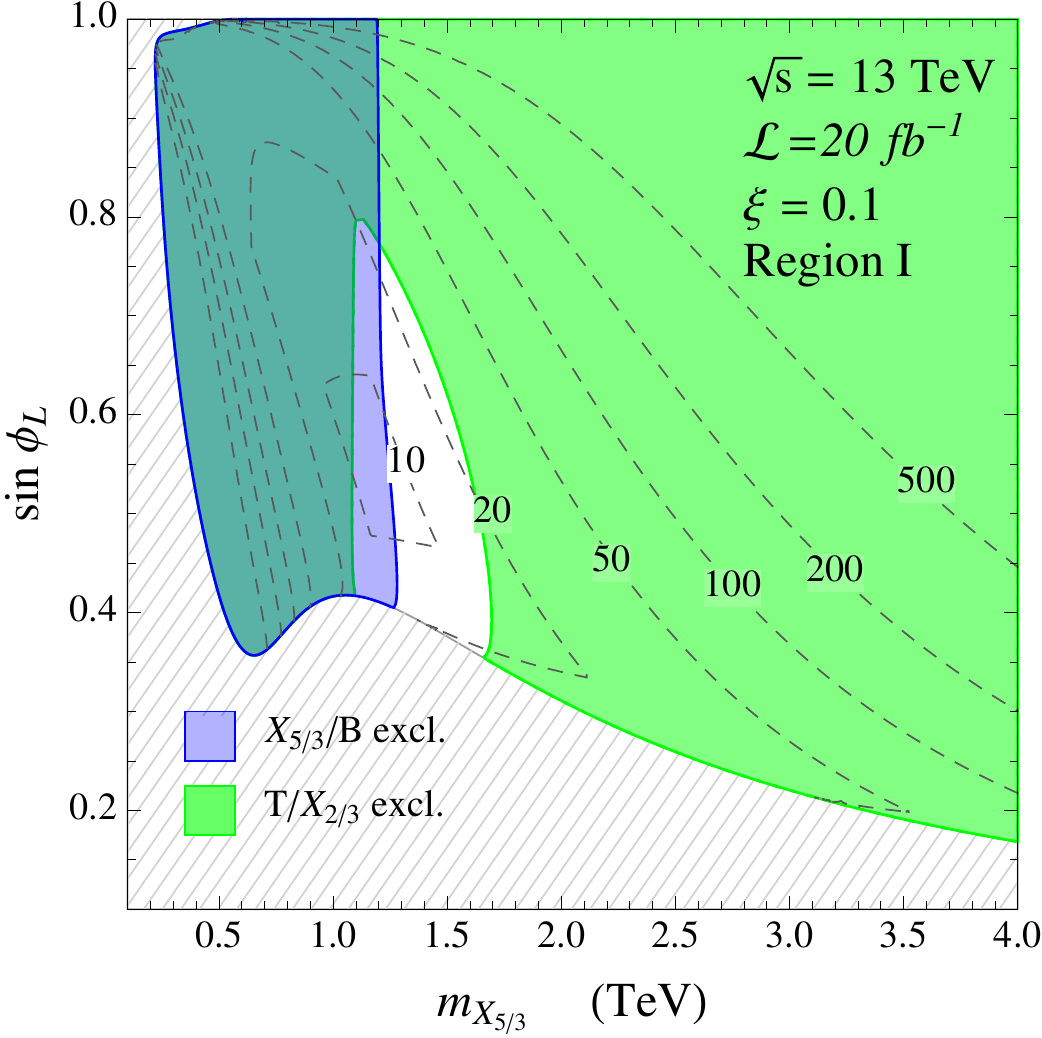}
\hspace{2em}
\includegraphics[width=.42\textwidth]{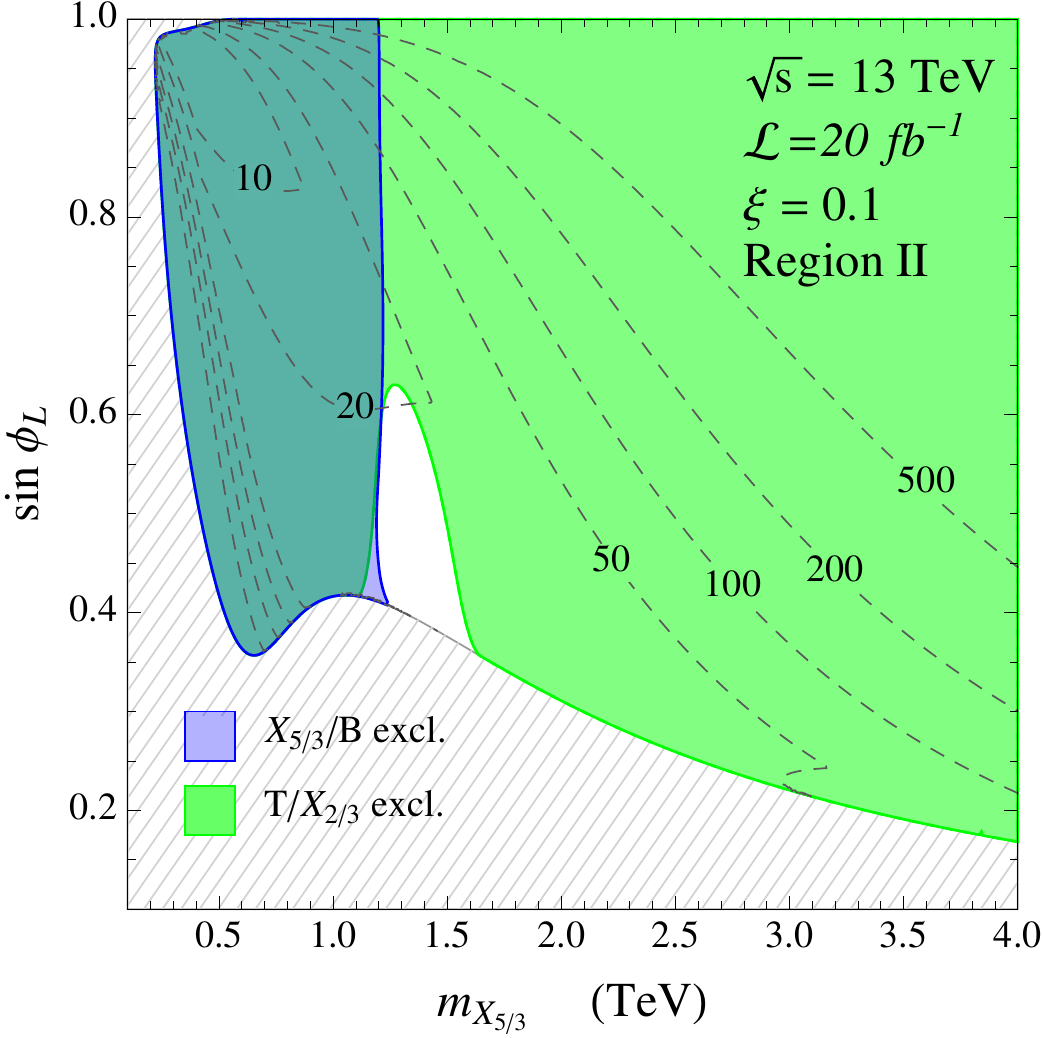}
\caption{Expected exclusion bounds in the $2$-site model with $\xi = 0.1$ and $c=0$ for the 13 TeV LHC run
with ${\mathcal L} = 20\ \mathrm{fb}^{-1}$ integrated luminosity. For further details see caption of fig.~\ref{fig:2-site_xi01_8TeV}.}
\label{fig:2-site_xi01_13TeV}
\end{figure}

As shown in fig.~\ref{fig:2-site_xi01_13TeV}, the 13 TeV LHC run with ${\cal L} = 20\ \mathrm{fb}^{-1}$ will be enough
to cover almost completely the $\xi = 0.1$ parameter space. In this case single production does not give a significant
improvement in the bounds for $c=0$. It can be checked that a mild improvement is instead expected for $c = \pm 1$.

\begin{figure}
\centering
\includegraphics[width=.42\textwidth]{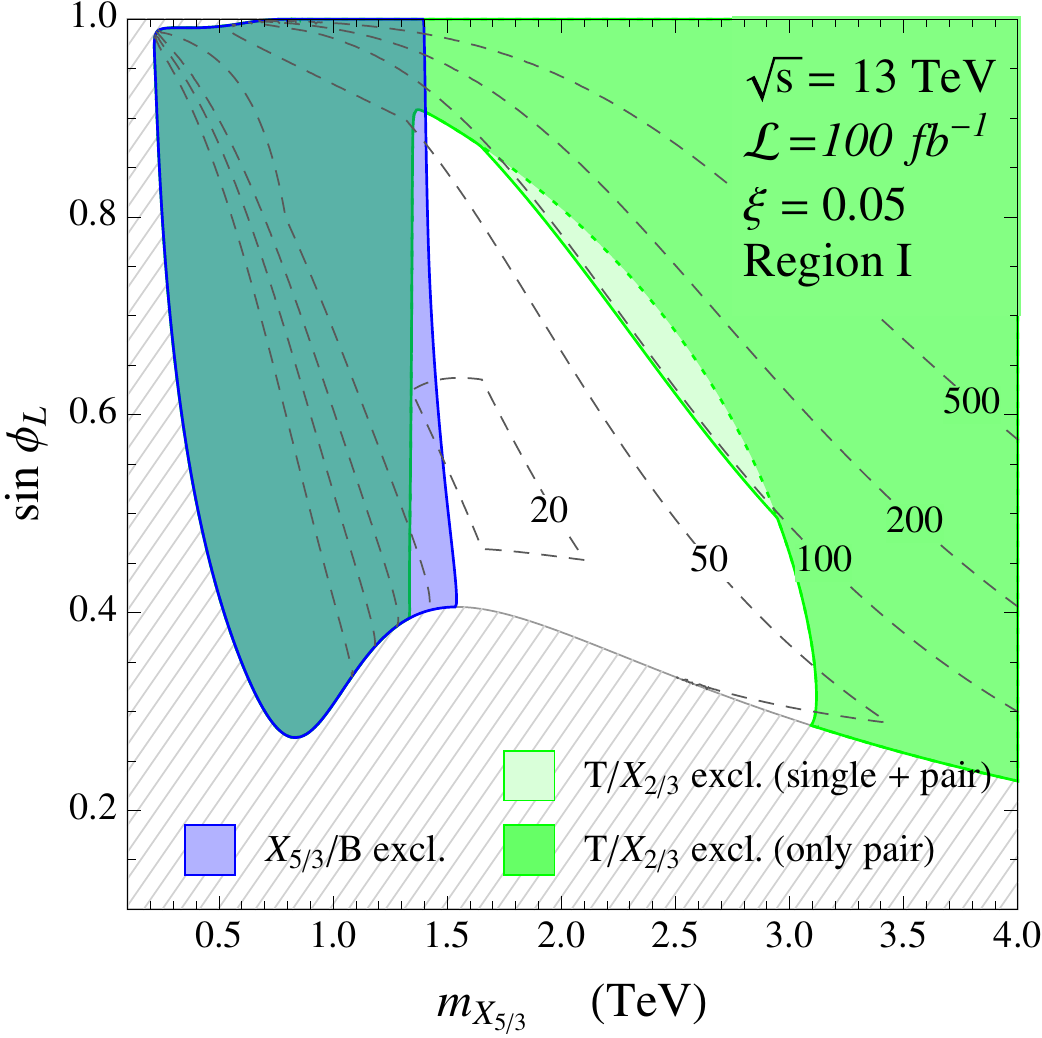}
\hspace{2em}
\includegraphics[width=.42\textwidth]{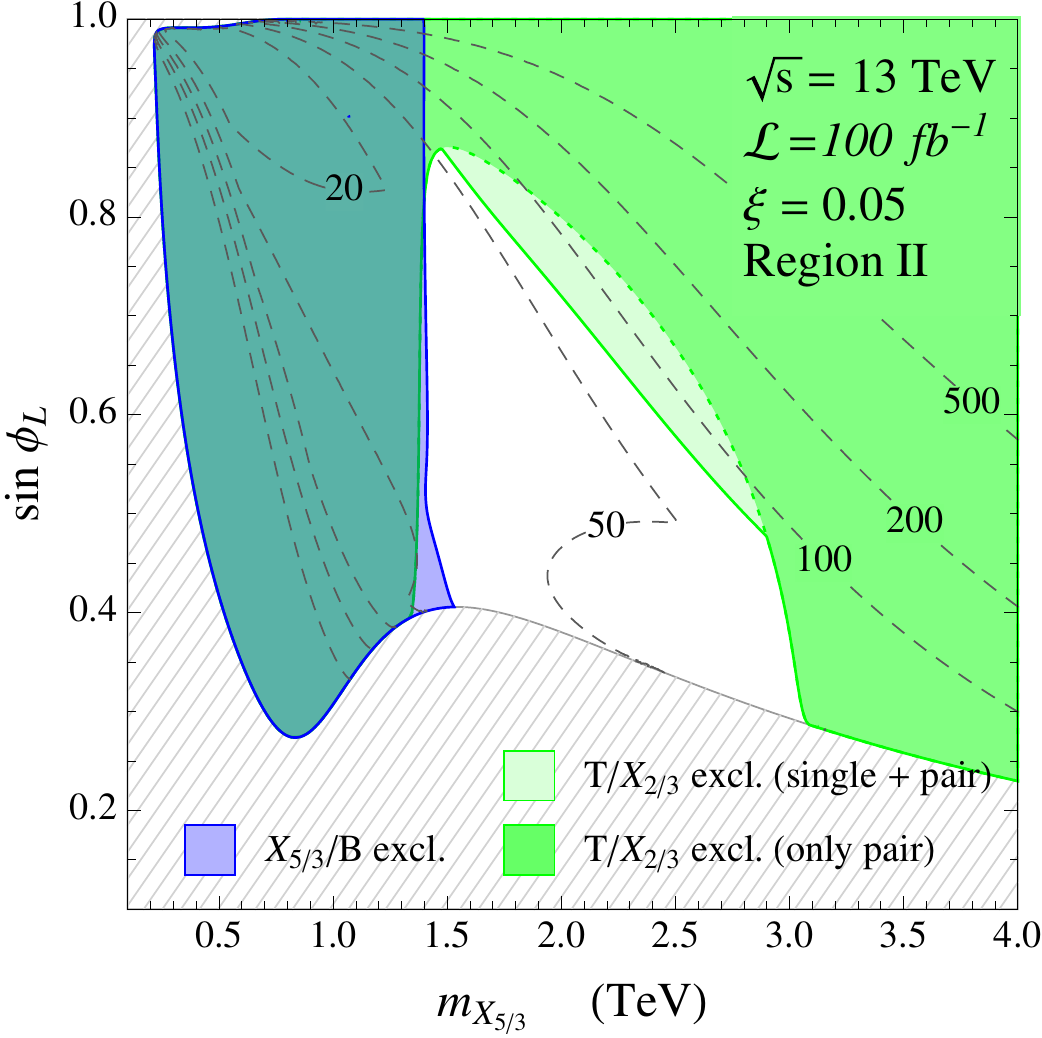}
\caption{Expected exclusion bounds in the $2$-site model with $\xi = 0.05$ and $c=0$ for the 13 TeV LHC data
with ${\cal L} = 100\ \mathrm{fb}^{-1}$ integrated luminosity.
For further details see caption of fig.~\ref{fig:2-site_xi01_8TeV}.}
\label{fig:2-site_xi005_13TeV}
\end{figure}

The expected 13 TeV LHC exclusion on the configurations with $\xi = 0.05$ and $c=0$ is presented
in fig.~\ref{fig:2-site_xi005_13TeV}. As for the $\xi = 0.1$ case, the addition of single-production searches for the charge-$2/3$
states can significantly improve the bounds, especially at relatively low integrated luminosity. For this value of $\xi$,
a significant part of the parameter space will still be allowed with ${\mathcal L} = 100\ \mathrm{fb}^{-1}$ integrated luminosity,
including configurations with minimal tuning $\Delta \sim 1/\xi \sim 20$.

\begin{figure}
\centering
\includegraphics[width=.5\textwidth]{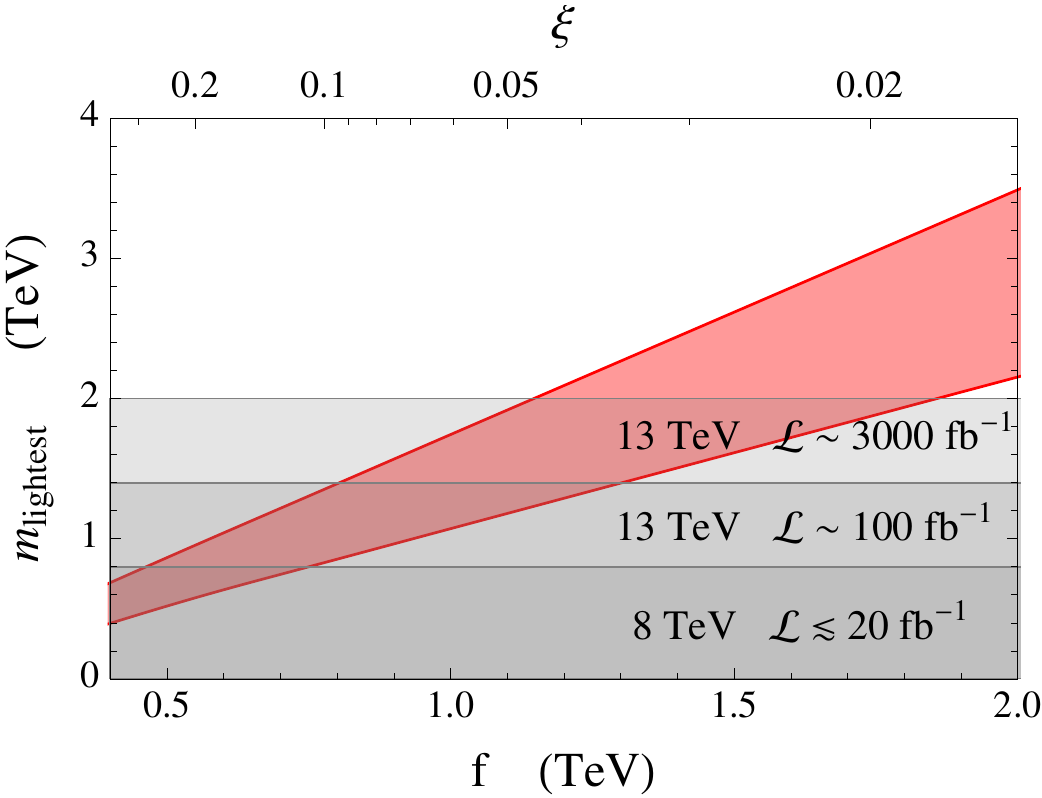}
\caption{Maximal value of the mass of the lightest fermionic partner in the $5 + 5$ $2$-site model as a
function of $f$. The red band is obtained by assuming that the relation in eq.~(\ref{eq:mhmt}) is valid with $20\%$ accuracy.
The gray bands correspond to the present and expected universal bounds coming from pair production searches.}
\label{fig:2-site_absolute}
\end{figure}

Finally in fig.~\ref{fig:2-site_absolute} we show the maximal value of the mass of the lightest top partner as a function
of $\xi$. This plot allows to translate the direct exclusion bounds into an upper bound on $\xi$. The current and
expected future pair production exclusions (denoted by the gray bands in the figure) show that at the end
of the LHC program configurations with $\xi \lesssim 0.05$ will be completely probed.

\section{Conclusions}

In this paper we introduced and analyzed some benchmark models for top partners in composite Higgs scenarios,
with the aim of visualizing in a concrete context the impact of the current exclusions and the expected reach of future searches.
We considered four simplified models in which only one light $\SO(4)$ multiplet of composite resonances is present.
The structure of the models is completely determined by the quantum numbers of the composite multiplet and does not rely on any extra
assumption. They are thus representative of a wide class of explicit models. In addition we also analyzed a more complete
$2$-site set-up in which, thanks to a collective breaking mechanism, the Higgs potential is partially calculable, thus providing
a link between the masses of the composite resonances, the Higgs mass and $\xi$.

The present bounds from the $8$~TeV LHC data mainly come from the QCD pair production channel and imply an absolute
lower bound on the mass of top partners $M_\Psi \gtrsim 800\ \mathrm{GeV}$. The inclusion of single production can
slightly improve the bound raising it to $M_\Psi \gtrsim 1\ \mathrm{TeV}$.
It must be noticed that, in the case of $\SO(4)$ singlet resonances, the size of the single production coupling is strongly related to the
size of the deviations in the $V_{tb}$ CKM matrix element. The region of the parameter space with sizable single production
can thus be also indirectly constrained from the measurements of $V_{tb}$. At present the indirect constraints
are dominant with respect to the direct LHC single production searches. In the $13$~TeV LHC run, instead, direct searches
are expected to have a better reach than indirect probes.
The $8$~TeV LHC bounds do not put a strong Naturalness pressure on the effective models, since configurations with small
tuning $\Delta \sim 10$ are still allowed.

In the case of no new-physics signal, the $13$~TeV LHC run is expected to substantially improve the bounds. The universal
constraint from pair production will exceed the $M_\Psi \simeq 1\ \mathrm{TeV}$ level in the first run 2 phase
(with an integrated luminosity ${\mathcal L} \simeq 20\ \mathrm{fb}^{-1}$) and could probe masses up to $M_\Psi \simeq 2\ \mathrm{TeV}$ at the end of the high-luminosity phase. Single production will also have a significant impact on the
exclusions allowing to test resonances with masses in the $M_\Psi \sim 3\ \mathrm{TeV}$ range.
In the light $\SO(4)$ fourplet scenario, configurations with small tuning $\Delta \lesssim 10$ will be completely tested
with ${\mathcal L} \sim 20\ \mathrm{fb}^{-1}$ integrated luminosity. Comparable exclusions for a light singlet will
instead require ${\mathcal L} \sim 100\ \mathrm{fb}^{-1}$.
The end of the LHC program, on the other hand, is expected to push the lower bound on the tuning
to the $\Delta \sim 20 - 50$ region.

Similar conclusions are found in the $2$-site scenario. In this case the bound on the resonance masses can be translated into
a lower bound on $\xi$. At present a sizable part of the parameter space with $\xi = 0.1$ is still allowed, including,
in particular, configurations with minimal tuning $\Delta \sim 1/\xi = 10$. All these configurations will be probed
at the $13$~TeV run with ${\mathcal L} \sim 20\ \mathrm{fb}^{-1}$. The high-luminosity LHC run will instead cover
all the configurations with $\xi \gtrsim 0.05$, implying a minimal amount of tuning $\Delta \gtrsim 20$.

In short, our conclusion is that current limits from the negative searches of top partners are not strong enough to put the idea
of a reasonably ``Natural'' composite Higgs in trouble. Parameter space regions with $\Delta\lesssim 10$ are allowed in all models
and this level of tuning is comparable with the one that is implied, in a rather model-independent way, by the present bounds
from Higgs coupling measurements. However while the latter is not expected to improve significantly in the next few years,
top partners direct searches will enormously extend their reach with the $13$~TeV LHC run.
We saw that a limited luminosity at  $13$~TeV, from around $20$~fb$^{-1}$ in the most favorable cases
to around $100$~fb$^{-1}$ in the less favorable ones, will be sufficient to probe levels of tuning deep in the $\Delta>10$ region.
If no significant excess will be seen, top partners direct searches will soon be singled out as the strongest bounds on the
composite Higgs scenario. 

If this will be the case, asking if and how plausibly ({\it{i.e.}}, at what price in terms of model-building complication)
the bound on $\Delta$ from negative top partners searches could be eluded will become a relevant question.
This could occur either if the top partners are light but evade the bounds because they are hard to detect
(see ref.~\cite{Serra:2015xfa} for a recent attempt), or if they are heavy but their mass, contrary to the generic
expectation outlined in the Introduction, is not directly linked to the level of tuning of the theory.
We do have examples of composite Higgs constructions, based on the so-called ``Twin Higgs''
mechanism~\cite{Chacko:2005pe,Chacko:2005un}, in which the latter option is
realized~\cite{Geller:2014kta,Barbieri:2015lqa,Low:2015nqa}.
Searching for alternatives to the twin Higgs mechanism, identifying the possible microscopic origin of the
twin Higgs symmetry and of its breaking and studying the phenomenological manifestation of these
scenarios~\cite{Craig:2015pha}, aside from heavy QCD- and EW-charged resonances beyond
the reach of the LHC, are topics that will be worth exploring.


\section*{Acknowledgments}

We thank A.~David and J.~Shu for useful discussions.
G.P. thanks CERN for hospitality during the completion of this work. The work of G.P. has been partially supported by
the Spanish Ministry MINECO under grants FPA2014-55613-P, FPA2013-44773-P, and FPA2011-25948,
by the Generalitat de Catalunya grant 2014-SGR-1450 and by the Severo Ochoa excellence program
of MINECO (grant SO-2012-0234).
A.W. acknowledges the MIUR-FIRB grant RBFR12H1MW and the ERC Advanced Grant no.~267985 (DaMeSyFla).


\appendix

\section{Derivation of the bounds}\label{sec:bounds}

In this appendix we briefly summarize the procedure used to derive the limits on the resonance messes.
For this task we recast the results of ref.~\cite{Matsedonskyi:2014mna}, with only minor differences
due to the use of the latest LHC data. The searches for exotic $X_{5/3}$ resonances and for charge $2/3$ states
are based on different decay channels, thus require slightly different approaches. We will consider them separately in
the following.

As a first case we consider the searches for the exotic $X_{5/3}$ states. The conservation of the electric charge forces
this resonance to decay exclusively through the channel $X_{5/3} \rightarrow W t$. Being part of an $\SO(4)$ $4$-plet,
the $X_{5/3}$ resonance has a leading coupling only to the $t_R$ component, while the coupling with the $t_L$
is generated only after EW symmetry breaking and is thus suppressed by additional $v/f$ factors.
The run 1 LHC searches exploit the pair production channel and are mainly focused on final states with a pair of same-sign
leptons. Currently the strongest bound $m_{X_{5/3}} > 880\ \mathrm{GeV}$ can be inferred from the CMS search
in ref.~\cite{Khachatryan:2015gza}. Although this search is focused on pair-produced fermionic resonances with charge $-1/3$,
it is expected to apply also for the $X_{5/3}$ state since both resonances lead to the same final states with somewhat
similar kinematics (see ref.~\cite{Mrazek:2009yu} for a more detailed discussion). For our purposes, however, the CMS
analysis of ref.~\cite{Khachatryan:2015gza} is too complex to be recast, since it combines several final states
(the most relevant being same-sign dileptons, leptons plus jets and multileptons) and relies on kinematic distributions.
A simpler search strategy, based on a cut-and-count analysis in same-sign lepton final states,
was instead used in some previous experimental works by CMS~\cite{cms53} and ATLAS~\cite{atlas051}
achieving an exclusion limit $m_{X_{5/3}} > 770\ \mathrm{GeV}$.\footnote{Notice that this limit is similar to the one obtained
in the recent CMS analysis by using exclusively the same-sign dilepton channel, as can be seen from
fig.~11 of ref.~\cite{Khachatryan:2015gza}.} These analyses can be more easily recast for our purposes as done in
ref.~\cite{Matsedonskyi:2014mna}. In particular the recast allows to straightforwardly take into account
the additional signal contributions coming from single production of the exotic $X_{5/3}$ resonance and from pair production
of the $B$ partner.\footnote{An alternative search strategy for the $X_{5/3}$ and $B$ resonances, which also focuses
on the single production channels, has been presented in ref.~\cite{Backovic:2014uma}.}

To estimate the future exclusion reach we instead started from the analysis of ref.~\cite{cms14}, which performs
a study of the $X_{5/3}$ searches in the dilepton channel at the $14\ \mathrm{TeV}$ LHC. In this case the single production
contribution to the signal has been estimated by assuming a reconstruction efficiency equal to $50\%$ of the
pair production one. This assumption is supported by the fact that the current ATLAS search in ref.~\cite{atlas051}
shows a similar relation between single and pair production efficiency~\cite{Matsedonskyi:2014mna}.

The exclusion bounds from our recast are summarized in fig.~3, 4 and 10 of ref.~\cite{Matsedonskyi:2014mna} and show how the
limits depend on the main single production coupling $W X_{5/3} t_R$. We used these results to derive the bounds
discussed in the main text.

Let us now discuss the analysis for the charge $2/3$ resonances. The current experimental exclusions obtained by the
ATLAS~\cite{Aad:2015kqa} and CMS~\cite{Chatrchyan:2013uxa} collaborations are only based on the pair production channel
and are reported as a function of the three branching ratios $BR(\widetilde T \rightarrow Wb)$, $BR(\widetilde T \rightarrow Zt)$
and $BR(\widetilde T \rightarrow ht)$.
The strongest exclusions come from the ATLAS analysis and range between
$730\ \mathrm{GeV}$ and $950\ \mathrm{GeV}$. In particular for a singlet resonance
($BR(\widetilde T \rightarrow Wb) \simeq 2 BR(\widetilde T \rightarrow Zt) \simeq 2 BR(\widetilde T \rightarrow ht) \simeq 1/2$)
the current bound is $m_{\widetilde T} > 790\ \mathrm{GeV}$ and is roughly comparable with the present bound
on the $X_{5/3}$ states.

The experimental analyses can not be easily adapted to the single-production channels and no estimate
of its impact on the exclusions has been presented so far by the experimental
collaborations.\footnote{An experimental study that includes the single production channel
has been presented in ref.~\cite{Aad:2015gdg}. However, the values of the single production coupling considered
in the analysis are too small to have an impact on the exclusions, so that the result can not be used to derive
a bound on the single production channel.}
A few theoretical analyses are however present in the literature. They mostly focus on single decay channels,
namely $\widetilde T \rightarrow Zt$~\cite{Basso:2014apa,Reuter:2014iya}, $\widetilde T \rightarrow ht$~\cite{Vignaroli:2012nf,Endo:2014bsa} and $\widetilde T \rightarrow Wb$~\cite{Ortiz:2014iza,Gripaios:2014pqa,Andreazza:2015bja},
with the exception of ref.~\cite{Li:2013xba} which considers a combination of $\widetilde T \rightarrow Wb$ and $\widetilde T \rightarrow ht$.

Following ref.~\cite{Matsedonskyi:2014mna}, for our analysis we performed a simple recast of the search in the
$Wb$ channel proposed in ref.~\cite{Ortiz:2014iza}. The $8\ \mathrm{TeV}$ limits have been straightforwardly
adapted by reconstructing the number of signal events required for the exclusion ($S_{exc} = 26$) and the cut efficiency.
The extension to the $13\ \mathrm{TeV}$ case, instead, has been done by naively assuming that $S_{exc}$ and the
cut efficiency are the same as the ones at $8\ \mathrm{TeV}$.

The results we obtained for the run 1 LHC exclusions are in fair agreement with the limits derived in ref.~\cite{Li:2013xba},
which also exploits the $Wb$ channel. Some discrepancy is instead present in the comparison
with ref.~\cite{Andreazza:2015bja}, whose bounds on the $W\widetilde Tb_L$ coupling as a function of the $\widetilde T$
mass are roughly a factor $2$ weaker. The bounds coming from the $Zt$ and $ht$ channels, on the other hand,
seem consistently weaker than the ones from the $Wb$ channel, although a significant spread in the results (especially
for large $\widetilde T$ masses) is present among the various estimates.


\end{document}